\documentclass[apj]{emulateapj}

\newcommand{\eg}{e.g.,}

\newcommand{\etal}{et~al.\null}

\newcommand{\nar}{\ref@jnl{New A Rev.}}%

\shorttitle{Searching for Ly$\alpha$ halos}
\shortauthors{Feldmeier et~al.}

\begin{document}

\title{Searching for Neutral Hydrogen Halos around $z\sim2.1$ and $z\sim3.1$ 
Ly$\alpha$ Emitting Galaxies}

\author{
John J. Feldmeier\altaffilmark{1,2},
Alex Hagen\altaffilmark{3,4},
Robin Ciardullo\altaffilmark{2,3,4},
Caryl Gronwall\altaffilmark{3,4},
Eric Gawiser\altaffilmark{2,5},
Lucia Guaita\altaffilmark{2,6},
Lea M. Z. Hagen\altaffilmark{3,4},
Nicholas A. Bond.\altaffilmark{7},
Viviana Acquaviva\altaffilmark{8},
Guillermo A. Blanc\altaffilmark{9},
Alvaro Orsi\altaffilmark{10,11},
Peter Kurczynski\altaffilmark{5}
}

\altaffiltext{1}{Department of Physics and Astronomy, Youngstown State University, 
Youngstown, OH 44555, USA; jjfeldmeier@ysu.edu}

\altaffiltext{2}{Visiting astronomer, Cerro Tololo Inter-American Observatory, National 
Optical Astronomy Observatory, which are operated by the Association of Universities for 
Research in Astronomy, under contract with the National Science Foundation.}

\altaffiltext{3}{Department of Astronomy \& Astrophysics, The Pennsylvania State University,
University Park, PA 16802, USA}

\altaffiltext{4}{Institute for Gravitation and the Cosmos, The Pennsylvania State University, University
Park, PA 16802, USA}

\altaffiltext{5}{Department of Physics and Astronomy, Rutgers, The State University of 
New Jersey, Piscataway, NJ 08854, USA}

\altaffiltext{6}{Oskar Klein Cosmology Centre, Department of Astronomy, Stockholm University, 
Stolkholm, Sweden}

\altaffiltext{7}{Cosmology Laboratory (Code 665), NASA Goddard Space Flight Center, Greenbelt, MD 20771, USA}

\altaffiltext{8}{Department of Physics, New York City College of Technology, 
City University of New York, 300 Jay Street, Brooklyn, NY 11201, USA}

\altaffiltext{9}{Observatories of the Carnegie Institution of Washington, 813 Santa Barbara Street, Pasadena, CA 91101, USA}

\altaffiltext{10}{Departamento de Astronom\'ia y Astrof\'isica, Pontificia Universidad Cat\'olica, Av. Vicu\~na Mackenna 4860, Santiago, Chile. }

\altaffiltext{11}{Centro de Astro-Ingenier\'ia, Pontificia Universidad Cat\'olica, Av. Vicu\~na Mackenna 4860, Santiago, Chile.}

\begin{abstract}
We search for evidence of diffuse Ly$\alpha$ emission from extended neutral hydrogen 
surrounding Ly$\alpha$ emitting galaxies (LAEs) using deep narrow-band images of the 
Extended Chandra Deep Field South.  By stacking the profiles of 187 LAEs at $z = 2.06$, 
241 LAEs at $z = 3.10$, and 179 LAEs at $z = 3.12$, and carefully performing low-surface 
brightness photometry, we obtain mean surface brightness maps that reach 9.9, 8.7, and 
$6.2 \times 10^{-19}$~ergs~cm$^{-2}$~s$^{-1}$~arcsec$^{-2}$ in the emission line.  
We undertake a thorough investigation of systematic uncertainties in our surface brightness 
measurements, and find that our limits are 5--10 times larger than would be expected from 
Poisson background fluctuations; these uncertainties are often underestimated in the literature.
At $z \sim 3.1$, we find evidence for extended halos with small scale lengths of
5--8 kpc in some, but not all of our sub-samples.  We demonstrate that sub-samples of 
LAEs with low equivalent widths and brighter continuum magnitudes are more likely to possess such halos.
At $z \sim 2.1$, we find no evidence of extended Ly$\alpha$ emission down to our
detection limits.  Through Monte-Carlo simulations, we also show that we would have detected 
large diffuse LAE halos if they were present in our data sets.  We compare these findings to 
other measurements in the literature, and discuss possible instrumental and astrophysical 
reasons for the discrepancies.
\end{abstract}


\section{Introduction}

The star formation rates of high and intermediate redshift galaxies depend on the complex
interplay between accretion and outflow.  While infalling gas from the cosmic web
provides fuel for starbursts \citep[\eg][]{dekel2009}, the winds of young stars, supernovae,
and active galactic nuclei cause feedback and dampen the intensity of the accretion
\citep[see][]{veilleux2005}.  The star formation history of a galaxy and its surrounding
circum-galactic medium (CGM) are thus inextricably intertwined, and any model which
seeks to reproduce the observed properties of these systems must take into account both of
these processes \citep[\eg][]{dave2011a, dave2011b}.

Observationally, there is strong evidence for  the presence of galactic-scale outflows around
luminous star-forming galaxies at high redshift.  At $z \sim 3$, the velocity offsets between 
resonance and non-resonance absorption features of Lyman-break galaxies (LBGs)
\citep[e.g.,][]{steidel1996, adelberger2003, shapley2003, steidel2010} and the relative strengths
and velocities of multiple-peaked profiles of Ly$\alpha$ in emission 
\citep[e.g.,][]{verhamme2006,tapken2007,verhamme2008,laursen2009,barnes2011,kulas2012}
all point to outflows being common in these high luminosity systems.  Outflows are also common in
lower-luminosity objects: the spectral analyses of \citet{mclinden2011} and \citet{berry2012} and 
more recently, \citet{hashimoto2013} clearly demonstrate that
Ly$\alpha$ emitting galaxies at $z \sim 2-3$ have strong galactic winds.  However,
while the evidence for outflows is compelling, there is as of yet no direct evidence
for the cold gas infall that is predicted by galaxy formation models \citep{steidel2010}.

Detecting the circum-galactic material around high-redshift galaxies is a difficult proposition. 
\citet{steidel2010} was able to infer the existence of this component out to $\sim 120$~kpc
by using angular pairs of unassociated $z \sim 2-3$ star-forming galaxies and looking for
Ly$\alpha$ and Ly$\beta$ in absorption.  However, a more direct approach  
is to search for circum-galactic material in emission.   Resonant scattering can cause the 
Ly$\alpha$ photons produced inside a galaxy to scatter hundreds or even thousands of 
times before escaping into intergalactic space.   Consequently, if a galaxy is surrounded by 
neutral hydrogen gas, that material is likely to be illuminated by Ly$\alpha$.  In fact, models 
by \citet{zheng2011} suggest that at $z = 5.7$, diffuse Ly$\alpha$ emission in the CGM
may be observable several hundred kpc from the site of its creation.  In addition to
inflows and outflows, a clumpy and inhomogeneous ISM may also create halo-like structures.       
\cite[\eg][]{verhamme2012}

The recent observations by \citet[][hereafter S11]{steidel2011} appear to confirm this
prediction.   By co-adding Keck, Hale, and Subaru narrow-band images of 92 LBGs
between $2.2 < z < 3.2$,  S11 claimed the detection of scattered Ly$\alpha$ from a diffuse 
halo that extends $\sim 80$~kpc from their mean composite galaxy.  The azimuthally-averaged 
surface brightness profile of this halo is exponential, with a scale length that is substantially 
larger than that of the galaxy's continuum, but similar to that measured for Ly$\alpha$ blobs 
\citep[\eg][]{steidel2000, matsuda2004,saito2006,prescott2012}.  This consistency led S11 to argue that Ly$\alpha$ halos 
are a generic feature of all high-redshift star-forming galaxies, and that these halos have
self-similar exponential profiles.

However, the galaxies targeted by S11 were all selected via the Lyman-break technique, and 
are thus amongst the brightest and most massive galaxies in the high redshift universe.
With stellar masses of  $\sim 2 \times 10^{10} M_{\odot}$, star formation rates (SFRs)
of $\sim 30 M_{\odot}$~yr$^{-1}$, and internal extinctions of $A_V \sim 1$ 
\citep[\eg][]{shapley2001, adelberger2005,stark2009,fink2010,labbe2010} these objects are in the extreme.  Most 
high-$z$-objects have much lower masses and star formation rates, and are not
detectable via the Lyman-break technique.  To test the ubiquity of Ly$\alpha$ halos, 
more representative samples of objects are needed.

Ly$\alpha$ emitting galaxies (LAEs) may be a better population for this purpose.  
Thousands of LAE candidates have been detected over a very large range of 
redshifts using a variety of observational techniques, and their properties have been
intensely studied \citep[\eg][and references within]
{ouchi2008,ouchi2010,rauch2008,fink2009b,wang2009,ostlin2009,nilsson2009,nilsson2011,
hu2010,ono2010,ota2010,hibon2010,tilvi2010,guaita2010,
cassata2011,cowie2011,adams2011,
ciardullo2012,barger2012,oteo2012,mallery2012,nakajima2012,krug2012}.  
Although LAEs can have Ly$\alpha$ luminosities similar to that of Lyman-break galaxies, most have
star formation rates (SFR $\sim 2 M_{\odot}$~yr$^{-1}$) and stellar masses 
($M \lesssim 10^{9} M_{\odot}$) that are an order of magnitude less \citep[\eg][]{gawiser2006b,pirzkal2007,lai2008,fink2009,ono2010,yuma2010}.  These are the
objects that will likely evolve into today $L^*$ galaxies like the 
Milky Way \citep[\eg][]{pirzkal2007,gawiser2007,guaita2010,salva2010,yajima2012}.

An additional advantage of LAEs are that they are relatively metal-poor and dust-poor, with most having 
internal extinctions $A_V \lesssim 0.5$ 
\citep[\eg][however also see \citet{fink2009} and \citet{nakajima2012} for contrasting views]{nilsson2007,gawiser2007, 
ono2010,acquaviva2011,fink2011a,fink2011b,blanc2011}.  
This lack of dust makes it more likely that the Ly$\alpha$ photons generated in the systems' 
star-forming regions will escape, although strong outflows in LBGs may also produce similar escape
fractions.  Finally, spectroscopy has shown that the contamination rate of photometrically-selected 
LAEs at $z < 3.1$ is small, less than 2\% \citep{gronwall2007,ouchi2008,berry2012}.  Thus, large samples 
of Ly$\alpha$ emitters are relatively straightforward to obtain.   

Recently, \citet[][hereafter M12]{matsuda2012} used the co-added images of 
$\sim 2100$ LAEs to conclude that the properties of Ly$\alpha$ halos depend on environment,
with the emission surrounding LAEs in low density environments being much fainter than
that for corresponding objects in overdense regions.   However, neither S11 or M12 
explored the systematic effects which limit the depth to which image stacking can be 
trusted.  Invariably, the true limit of any surface brightness analysis is significantly 
brighter than that which is inferred from simple counting statistics.

The S11 study focused on cluster LBGs at $2.3 < z < 3.1$ while M12 focused on LAEs at
$z\sim3.1$ in a range of galaxy environments.   In this paper, we investigate the extended
Ly$\alpha$ halos of field LAEs at similar redshifts.   In \S 2, we define our samples using the 
$z \sim 3.1$ observations of \citet{gronwall2007} and \citet{ciardullo2012}, and the $z \sim 2.1$ 
data of \citet{guaita2010}.   In \S 3, we describe our stacking technique and present the initial analysis of
the Ly$\alpha$ profiles of our LAE sub-samples.  Then in \S 4, we perform a detailed analysis of the 
various effects which limit the precision of stacking analyses, and show that the 
systematic uncertainties in this procedure have generally been 
underestimated.  In \S 5, we present a more detailed stacking 
analysis,
and demonstrate that there is evidence for extended Ly$\alpha$-emission in some,
but not all of our samples, 
down to surface brightnesses of $\approx 8 \times
10^{-19}$~ergs~cm$^{-2}$~s$^{-1}$~arcsec$^{-2}$ at $z \sim 3$ and $\approx 1\times
10^{-18}$~ergs~cm$^{-2}$~s$^{-1}$~arcsec$^{-2}$ at $z \sim 2$.  
We show that stacks containing sources that have relatively low Ly$\alpha$
equivalent widths and/or bright continuum magnitudes are more likely to show
extended Ly$\alpha$ emission than similar stacks with low 
continuum luminosities or high equivalent widths.  In \S 6, we compare our results
to previous studies via a series of Monte Carlo simulations, and demonstrate that halos 
such as those found by S11 would have easily been detected in our survey.  
In \S 7, we discuss possible explanations for our results, including
the extended telescope point-spread-function, the limitations of sky subtraction, the
intrinsic differences between galaxy populations, the effects of environment and orientation,
and the possible evolution of Ly$\alpha$ halos between $z \sim 3$ and $z \sim 2$. 
Finally, in \S 8, we summarize our results and discuss the possible ways of improving 
the constraints on Ly$\alpha$ halos with upcoming surveys and new Ly$\alpha$ 
radiative transfer models.

For this paper, we assume a $\Lambda$CDM cosmology with $\Omega_{\lambda}=0.70$, 
$\Omega_m=0.3$ and $H_0=70$~km~s$^{-1}$~Mpc$^{-1}$.  With these values, 
$1\arcsec = 8.32$~kpc at $z=2.1$ and $7.63$~kpc at $z=3.1$.

\section{The Sample of Ly$\alpha$ Emitters}

In order to test for the existence of diffuse emission around Ly$\alpha$ emitting galaxies, 
we consider three statistically complete sets of LAEs.  Each set is located in the 
Extended Chandra Deep Field South \citep[ECDF-S;][]{lehmer2005}, each was identified
via narrow-band imaging with the Mosaic II CCD camera \citep{muller1998} on the CTIO 
Blanco 4-m telescope, and each defines an LAE as an emission-line galaxy with a 
rest-frame Ly$\alpha$ equivalent width greater than 20~\AA.  The statistically complete 
sample of LAEs in each data set was defined by finding all LAEs with fluxes greater than 
the 90\% completeness level, where the completeness curve was determined from artificial 
star simulations.  

Our first set of LAEs is that found by \citet{gronwall2007} using a 50~\AA\ 
full-width-half-maximum (FWHM) filter centered 4990~\AA\null.  This dataset, hereafter 
called C-O3, covers the redshift range $3.08 < z < 3.12$ and is 90\% complete to a monochromatic 
Ly$\alpha$ emission-line luminosity limit of $\log L = 42.1$~ergs~s$^{-1}$.  A total of 254 
objects were identified in this survey, though only 156 are members of the statistically 
complete sample.  To date, 67 of these LAEs have been spectroscopically
confirmed as $z \sim 3.1$ star-forming galaxies \citep[the first 61 confirmations 
are reported in][]{gawiser2007}.  The median seeing for this survey was $1\farcs0$. 

The second set of LAEs is that found by \citet{ciardullo2012}, using a 57~\AA\ FWHM filter 
centered at 5010~\AA\null.  This LAE catalog, hereafter called K-O3, is 90\% complete to 
$\log L > 42.3$~ergs~s$^{-1}$, and has considerable overlap with that of  \citet{gronwall2007}.  However, 
because the K-O3 central wavelength is slightly redder than that of C-O3, the dataset also
contains 68 new objects in the redshift range $3.10 < z < 3.15$.   This K-O3 survey 
discovered 199 LAE candidates; 130 of these are in the statistically complete sample.
The median seeing for this survey was $1\farcs1$.

Our third sample of LAEs examines a different redshift regime.  By using a 50~\AA\ FWHM 
filter centered at 3727~\AA, \citet{guaita2010} were able to identify 250~LAE candidates in 
the redshift range $2.04 < z < 2.09$.  \citet{guaita2011} further refined this sample, 
by improving the photometry, removing one probable AGN, and using 
{\sl Hubble Space Telescope\/} morphological measurements to exclude 34 likely interlopers.  This left  
a final sample consisting of 201 objects, of which 73 are brighter than a 90\% completeness limit of 
$\log L = 41.8$~ergs~s$^{-1}$.  We define this as the O2 LAE sample.  The median seeing 
for this survey was $1\farcs4$.

Since all three surveys targeted a region of space with deep Chandra data, most quasars and 
other active galactic nuclei have been eliminated from the samples based on their X-ray
emission \citep{lehmer2005, virani2006, luo2008}.  Similarly, because the ECDF-S has been 
subjected to deep near-UV (1700--2800\AA\null) surveys by the {\sl GALEX} and {\sl Swift} satellites \citep{schim2003, 
hoversten2009}, 
most foreground interlopers have already been identified.  (For the redshifts 
under consideration, the rest-frame near-UV lies blueward of the Lyman break.)  Thus, our samples of 
LAEs  should be relatively  clean.  Of the 73 $z \sim 3.1$~LAEs with spectroscopic confirmations, 
{\it all\/} have redshifts that place the galaxy between $3.08 < z < 3.12$.  Similarly, of 
the 13 $z = 2.1$ objects with spectroscopic redshifts \citep{berry2012}, there are no foreground contaminants.

Finally, we should note that almost 300 LAEs in the E-CDFS (171 at $z \sim 3.1$
and 108 at $z \sim 2.1$) have been imaged in the rest-frame ultraviolet with 
{\sl Hubble Space Telescope's Advanced Camera for Surveys\/}.  An analysis of these 
data by \citet{bond2012} has shown that in $V_{606}$, LAEs in the redshift
range $2 < z < 3$ are extremely compact, with median half-light radii less than 
$0\farcs 2$. In fact, at $z = 3.1$, the largest half-light radius recorded is 
$0\farcs 29$ and the largest half-light radius at $z = 2.1$ 
is $0\farcs 43$.  In both cases, this is much smaller than the seeing of the 
ground-based images under consideration ($1\farcs 0$ and $1\farcs 4$), respectively.  
Similar results have been found for other LAE samples over a wide range of redshift 
\citep[\eg][and references within]{malhotra2012}.  Consequently, if the Ly$\alpha$ 
emission emanating from these objects has the same spatial 
distribution as the UV light, the profiles of these objects should be consistent
with those of the frames' point spread functions.

\section{Initial Stacking of Ly$\alpha$ Emitters}

To search for extended Ly$\alpha$ emission, we began with the stacked Mosaic~II
images obtained by \citet{gronwall2007}, \citet{guaita2010}, and \citet{ciardullo2012}.
First, a small $27\arcsec \times 27\arcsec$ 
(101 $\times$ 101 pixel) region was extracted about the center of each LAE\null.  To avoid
edge effects, any source close to the CCD boundaries was excluded at this step.  This left
us with 146, 124, and 69 sources in the statistically complete C-O3, K-O3, and O2 samples, and 
241, 179, and 187 objects in the total samples of these surveys.  
Next, to examine the behavior of Ly$\alpha$ halos in our highest mass objects, 
we used the broadband magnitudes of the MUSYC survey \citep{gawiser2006a} to define 
sub-samples of LAEs more luminous than $R_{AB} = 25.5$.  With this limit, these
LAE sub-samples have photometric and spectroscopic properties similar to those of
the ``LAE only'' sample of LBGs studied by S11\null.  These sub-samples, 
which we will refer to as ``UV bright,'' contain 28, 19, and 27 sources 
in the C-O3, K-O3, and O2 samples, respectively.  

Finally, to examine whether equivalent width has any affect on the properties of Ly$\alpha$
halos, we ranked our LAEs by their (photometrically-derived) rest-frame equivalent widths, and created
sub-samples of objects in the highest and lowest quartiles of the distribution.  
We refer to these sub-samples as  ``high EW'' and ``low EW,'' and each contains
60, 44, and 50 objects in the C-O3, K-O3 and O2 datasets, respectively.   
For the low EW sub-samples, the rest-frame equivalent widths range from
20~\AA\ to 60~\AA; for the high-EW sub-set, EW$_0 > 74$~\AA, with
most of the objects having rest-frame equivalent widths greater than 100~\AA\null.
We note that since the distribution of LAE
equivalent widths has (at best) a weak dependence on galaxy luminosity \citep{gronwall2007,ciardullo2012}, 
samples chosen by equivalent width and UV-brightness are somewhat orthogonal.  For example, in the 
C-O3 data, only $\sim 25\%$ of the low-equivalent width LAEs (16 out of 60) are members of the 
UV-bright sample; for the K-O3 sample, this fraction is $\sim 35\%$ (16 out of 44), and 
for O2, the fraction is $\sim 44\%$ (22 out of 50).
A summary of the different sub-samples, and other relevant information, is given in Table~\ref{table:samples}.

\begin{deluxetable*}{lcccccc}
\tablewidth{0pt}
\tablenum{1}
\tabletypesize{\scriptsize}
\tablecaption{Sub-Sample Properties\label{table:samples}}
\tablehead{
\colhead{Sample}
& \colhead {Number of LAEs}
& \colhead {$\langle z \rangle$}
& \colhead {Median Seeing}
& \colhead {Effective Narrow-band}
& \colhead {EW\tablenotemark{a}}
& \colhead {EW\tablenotemark{a}}\\
& \colhead {} 
& \colhead {} 
& \colhead {} 
& \colhead {Exposure Time (hrs)}
& \colhead {Range (\AA)}
& \colhead {Median (\AA)}  
}
\startdata
C-O3 Total                   & 241 & 3.10 & $1\farcs0$ & 24.00 & 20 - 366 &  54\\ 
C-O3 Statistically Complete  & 146 & 3.10 & $1\farcs0$ & 24.00 & 20 - 366 &  51\\
C-O3 UV Bright               &  28 & 3.10 & $1\farcs0$ & 24.00 & 20 - 158 &  30\\
C-O3 High EW                 &  60 & 3.10 & $1\farcs0$ & 24.00 & 82 - 366 & 103\\
C-O3 Low EW                  &  60 & 3.10 & $1\farcs0$ & 24.00 & 20 -  36 &  27\\
K-O3 Total                   & 179 & 3.12 & $1\farcs1$ & 15.67 & 20 - 628 &  63\\
K-O3 Statistically Complete  & 124 & 3.12 & $1\farcs1$ & 15.67 & 20 - 628 &  60\\
K-O3 UV Bright               &  19 & 3.12 & $1\farcs1$ & 15.67 & 22 -  49 &  23\\
K-O3 High EW                 &  44 & 3.12 & $1\farcs1$ & 15.67 & 94 - 628 & 125\\
K-O3 Low EW                  &  44 & 3.12 & $1\farcs1$ & 15.67 & 20 -  40 &  29\\
O2 Total                     & 187 & 2.07 & $1\farcs4$ & 35.75 & 20 - 933 &  43\\
O2 Statistically Complete    &  69 & 2.07 & $1\farcs4$ & 35.75 & 27 - 933 &  70\\
O2 UV Bright                 &  27 & 2.07 & $1\farcs4$ & 35.75 & 20 -  60 &  26\\ 
O2 High EW                   &  46 & 2.07 & $1\farcs4$ & 35.75 & 74 - 933 &  98\\
O2 Low EW                    &  46 & 2.07 & $1\farcs4$ & 35.75 & 20 -  28 &  23\\
\enddata
\tablenotetext{a}{Equivalent widths are all rest-frame, and were determined from photometry.}
\end{deluxetable*}

After defining the LAE sub-samples, the next step involved co-adding the individual
LAE images to create a high signal-to-noise stacked image.  We did this by
scaling each object to a common narrow-band flux level and then combining
the images, using the \textsf{imcombine} task within 
IRAF\null \footnotemark \footnotetext{IRAF is
distributed by the National Optical Astronomy Observatory, which is operated by the
Association of Universities for Research in Astronomy (AURA) under cooperative agreement
with the National Science Foundation.}, and adopting a ($3 \, \sigma$) sigma-clipping 
algorithm with a scheme which weighted each source by its total flux.  
This procedure is different than the stacking procedure of S11, who used straight averages with
masking for their stacks, and is slightly different than M12, who used a median combination, but
did not scale or weight their sources.  
To avoid any effects of interpolation, we used only integer pixel shifts for each stack; 
this procedure has the effect of slightly increasing the radial surface brightness profile.
The stacking procedure was used both on the original 
narrow-band images themselves, and on continuum-subtracted images, where the flux level 
at 5000~\AA\ was defined 
using the $B$ and $V$ frames of the MUSYC survey \citep{gawiser2006a}.  For the O2 data,   
the continuum flux at 3727~\AA\ was derived from publicly available $U$ and $B$ frames taken with 
the Wide Field Imager (WFI) of the ESO 2.2 m telescope \citep{hildebrandt2006} that
were reprojected to match the MUSYC BVR image of the same field.  These 
continuum subtracted images have the advantage of decreasing the contamination from 
nearby continuum sources, but come with the penalty of decreased signal-to-noise and possible 
mis-matches in the frames' point-spread-functions (PSFs).    

Finally, as a control on our experiments, we also analyzed a catalog of point sources 
initially found by \citet{altmann2006}  on the MUSYC broad-band images.  This catalog,
which consists of objects with the colors of Galactic K- and M-type stars, was
further refined by \citet{bond2011} using {\sl HST-WFC3} Early Release Science images
\citep[ERS;][]{windhorst2011} of the GOODS-S region.  These objects were re-identified on our
narrow-band frames, and stacked in a manner identical to that for the LAEs.  Fifty-five
stars formed the point-source stack on our C-O3 and K-O3 images; for the O2 data, our 
control image was created using the median of 19 stars.

\begin{figure*}
\figurenum{1}
\plotone{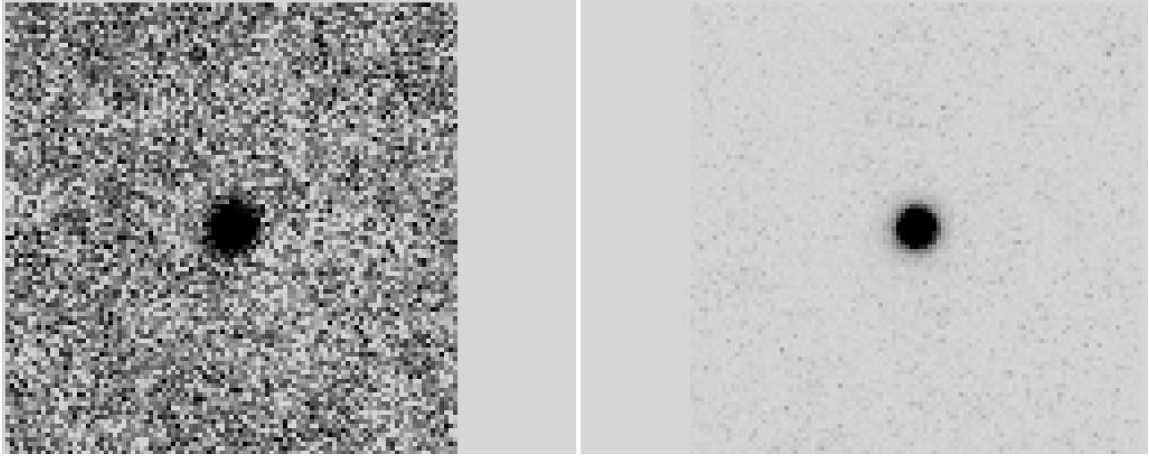}
\caption[Comparisons of Stacking]{\label{fig:imagecomp} 
On the left is the median-combined statistically complete sub-sample of 146 C-O3 Ly$\alpha$ emitters;
on the right is the median-combined sample of 55 stellar sources from the same image.  The greyscale
stretch is linear for each stack, and ranges from -1$\sigma$ to 0.15$p$, where $\sigma$ is the standard deviation
of the pixels near the corners of each image and $p$ is the peak pixel value in ADU.  
There is no apparent difference between the two stacks.  Formally, the FWHM of the Ly$\alpha$ 
emitters is 8\% larger than the stellar sources, but the uncertainties are of 
comparable magnitude (5\%).}
\end{figure*}

Figure~\ref{fig:imagecomp}  compares the median-stacked image of the 146 $z=3.1$ LAEs in the
statistically complete C-O3 sub-sample to that of the field's point sources.  
A careful examination of the figure reveals no compelling evidence for extended Ly$\alpha$
emission.  The results for the other stacks are similar: there is no clear
signature of extended Ly$\alpha$ emission in any of the LAE sub-samples.
Table~2 quantifies this result by listing the FWHM of a two-dimensional
Gaussian fit to each image stack, as measured with the {\tt fitpsf} task within IRAF\null.
The uncertainties on the fits for each stack are derived from a bootstrap analysis, in which 
we randomly resampled the sources (with replacement) and repeated the 
{\tt imcombine} and {\tt fitpsf} steps 30 times for each stack.  For comparison, the FWHM
of the point-source stacks are also shown, and the LAE measurements are
tabulated with respect to these values.  As the table demonstrates, 
there is a suggestion that some of the LAE stacks have broader radial profiles than 
that of the PSF stars.  However, the uncertainties in this type of analysis are large
and no obvious trends are apparent between the different sub-samples.
Moreover, this simple procedure does not fully account for the effects of 
centroiding errors: because the LAEs are much fainter than the comparison stars,
the FWHM of their stacks are expected to be wider due to positional uncertainties.  
A more sophisticated analysis is therefore needed to find evidence of Ly$\alpha$ 
halos.

\begin{deluxetable*}{lcccccc}
\tablewidth{0pt}
\tablenum{2}
\tabletypesize{\scriptsize}
\tablecaption{Results from Two-Dimensional Gaussian fitting\label{table:fitpsf}}
\tablehead{
&\multicolumn{2}{c}{C-O3} &\multicolumn{2}{c}{K-O3} &\multicolumn{2}{c}{O2} \\
\colhead{Sample} &\colhead{FWHM} &\colhead{Ratio\tablenotemark{a}}
&\colhead{FWHM} &\colhead{Ratio\tablenotemark{a}}
&\colhead{FWHM} &\colhead{Ratio\tablenotemark{a}} \\
\colhead{} & \colhead {(arcsec)}  & \colhead{} & \colhead{(arcsec)} & \colhead{} & \colhead{(arcsec)} & \colhead{}}
\startdata
Stacked Stars                     &  1.23 $\pm$ 0.03 & 1               & 1.23 $\pm$ 0.04 & 1               & 1.40 $\pm$ 0.09 & 1               \\
Total Sample                      &  1.39 $\pm$ 0.04 & 1.13 $\pm$ 0.04 & 1.28 $\pm$ 0.03 & 1.04 $\pm$ 0.04 & 1.53 $\pm$ 0.07 & 1.09 $\pm$ 0.09 \\
Statistically Complete            &  1.33 $\pm$ 0.05 & 1.08 $\pm$ 0.05 & 1.32 $\pm$ 0.02 & 1.07 $\pm$ 0.03 & 1.58 $\pm$ 0.07 & 1.13 $\pm$ 0.09 \\
UV Bright                         &  1.36 $\pm$ 0.07 & 1.11 $\pm$ 0.06 & 1.25 $\pm$ 0.05 & 1.01 $\pm$ 0.05 & 1.54 $\pm$ 0.17 & 1.10 $\pm$ 0.14 \\
High Equivalent Width             &  1.19 $\pm$ 0.07 & 0.97 $\pm$ 0.06 & 1.19 $\pm$ 0.04 & 0.96 $\pm$ 0.04 & 1.48 $\pm$ 0.07 & 1.06 $\pm$ 0.08 \\
Low Equivalent Width              &  1.35 $\pm$ 0.09 & 1.10 $\pm$ 0.08 & 1.28 $\pm$ 0.05 & 1.04 $\pm$ 0.04 & 1.31 $\pm$ 0.08 & 0.94 $\pm$ 0.08 \\
Difference Total                  &  1.28 $\pm$ 0.04 & 1.04 $\pm$ 0.04 & 1.21 $\pm$ 0.23 & 0.99 $\pm$ 0.19 & 1.09 $\pm$ 0.11 & 0.78 $\pm$ 0.09 \\
Difference Statistically Complete &  1.45 $\pm$ 0.08 & 1.18 $\pm$ 0.07 & 1.45 $\pm$ 0.18 & 1.18 $\pm$ 0.15 & 1.62 $\pm$ 0.15 & 1.16 $\pm$ 0.13 \\
\enddata
\tablenotetext{a}{Defined as FWHM (sub-sample) / FWHM (stacked stars), with the uncertainties from each sample added in quadrature.}
\end{deluxetable*}

\section{Surface Brightness Limitations}

Before proceeding further, we need to examine the realistic limits of our surface
brightness measurements, and properly characterize any systematic uncertainties.  This is not a 
straightforward process.  Surface photometry at faint flux levels has additional 
uncertainties above and beyond those produced by Poisson statistics.  Large-scale 
flat-fielding errors, the structure of the PSF at large radius, and errors in the estimation 
of the sky background all introduce systematic uncertainties into surface brightness measurements, 
and these errors dominate the error budget at low surface brightnesses.  Such uncertainties 
are well known in the astronomical literature, \citep[\eg][]{morrison1994, zheng1999, gonzalez2005, zibetti2005, krick2006,
bernstein2007} and counter-measures involving better telescope baffling \citep[\eg][]{grundahl1996}, advanced flat-fielding
techniques (such as drift-scanning, or creating high quality dark-sky flats; \eg~Zheng \etal 1999, Gonzalez et al. 2005), 
and using optical systems that minimize internal reflections \citep{slater2009} have been developed.  There has also been numerous 
efforts to develop more robust and precise algorithms for flat-fielding, sky subtraction, 
and the removal of foreground objects in ultra-deep surface photometry studies 
\citep[\eg][]{morrison1994,melnick1999,gonzalez2000,krick2006,mart2008,paudel2013}.  Unfortunately, these techniques 
are not always applied in practice or are logistically difficult to implement.  As a result, surface photometry 
on large telescopes with wide-field imagers will have a non-neglible amount of systematic uncertainty at low
surface brightness.  This applies to our own dataset, but it is very likely that other observations 
with similar systems will also have these effects.

\begin{figure*}
\figurenum{2}
\plotone{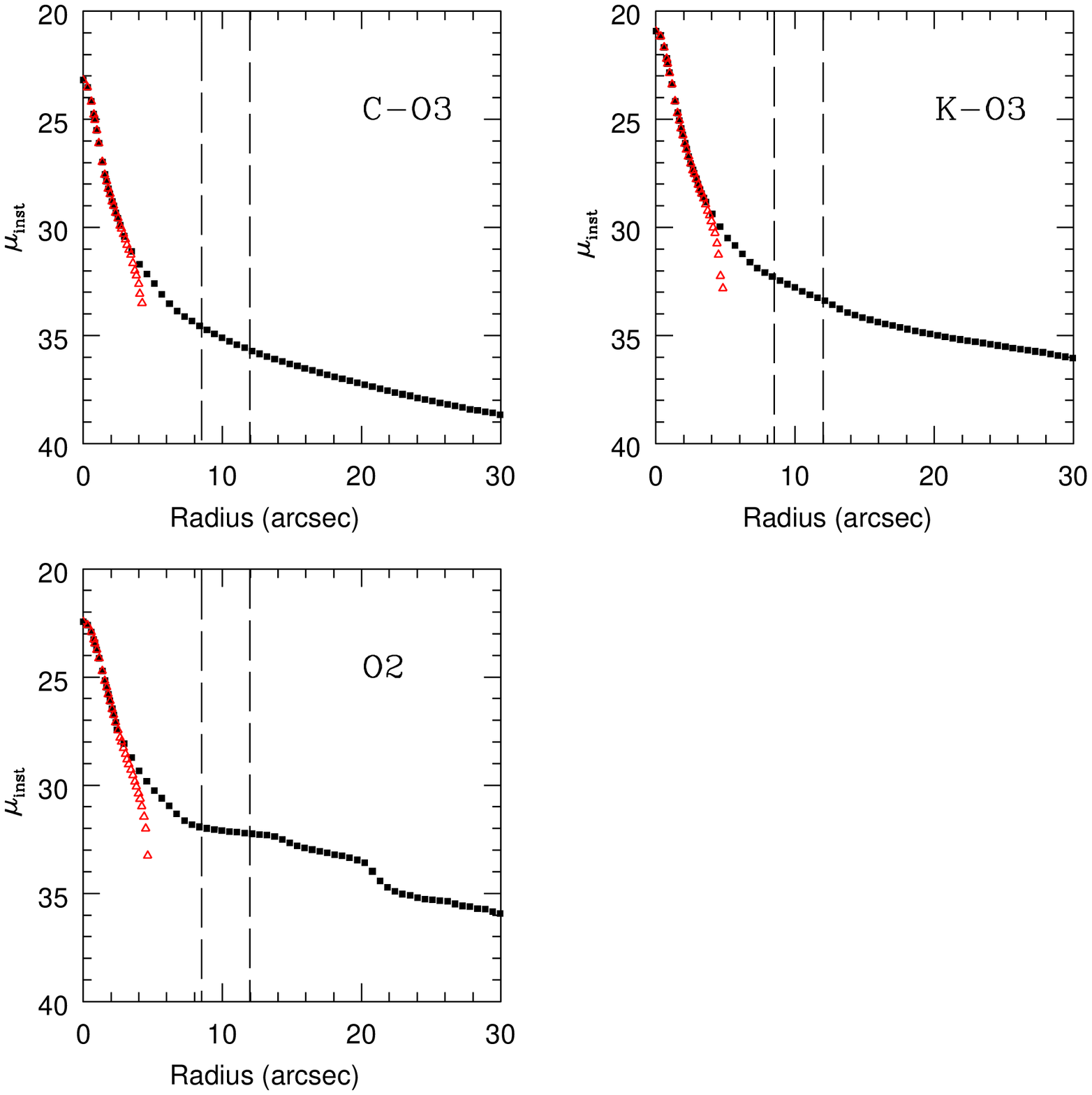}
\caption[ Large-Scale PSF]{\label{fig:psf} 
The azimuthally averaged surface brightness profile of a star constructed for each of our 
narrow-band images.  The instrumental surface brightness scale, $\mu_{inst}$ is 
defined as 1~ADU~$s^{-1}$ = 25 mag~arcsec$^{-2}$.
The red open triangles denote the small-scale PSF, as measured by bright 
stars and the DAOPHOT software package.  The filled squares represent the merger of the 
DAOPHOT PSF (at small radii) with the azimuthally averaged profile of a saturated star (at 
large radii), where the large-scale PSF was normalized to the small-scale PSF.  For reference, 
the dashed vertical lines shows the radial extent studied by S11 ($12\arcsec$) and M12 ($8\farcs5$) in their analyses.
}
\end{figure*}

The dominant systematic uncertainty for ultradeep surface photometry is usually the large-scale
flat-fielding error \citep{morrison1994,feldmeier2004}.  In order to empirically determine
this error on the C-O3, K-O3, and O2 survey data, we followed the procedures described in 
Feldmeier \etal~(2002, 2004) and \citet{rudick2010}.   In brief, we began by using DAOPHOT 
\citep{stetson1987} to find and measure the magnitude of every point source on the
frame.  We then determined a more precise small-scale PSF for each image (out to the larger 
radius of $\sim 5\arcsec$), while masking out all of the point sources down to an estimated level of 
$0.1/t_{\rm exp}$~ADU, where $t_{\rm exp}$ is the exposure time and ADU is the number of counts
above the sky value.  We next identified a 
saturated star in each image, and, after masking all other nearby sources, used it to
construct an azimuthally averaged PSF out to a much larger radius ($50\arcsec$ to $60\arcsec$,
depending on the image and the brightness of the chosen star).  As was found originally 
by \citet{king1971}, the PSF of a ground-based telescope extends out to a very large 
radius, and has a dramatic change in slope in the far field.  As displayed in Figure~\ref{fig:psf}
once outside of $\sim 4\arcsec$ (i.e., $\sim 30$~kpc at the redshift of our objects), 
the PSF contains flux far above that which would be predicted from a simple Gaussian
extrapolation.  Note the difference between the PSF of
the O2 image and that of the two images near 5000~\AA.  This is due to the differing 
patterns of internal reflections between the CCD dewar window and the narrow-band filter
\citep{slater2009}.  Measurements of the large-scale PSF by other authors 
\citep[see the detailed discussion by][]{bernstein2007} indicate that the shape of this 
function depends not only on the telescope, instrument, and wavelength, but on the observing 
season as well, as observations taken several months apart may have very different 
power-law-slopes at large radii.  Consequently, any analysis which relies on 
subtracting the large-scale PSFs of different telescopes and/or instruments must be 
treated with suspicion.  To be conservative, the remainder of this paper will focus 
solely our our narrow-band data alone; no attempt will be made to interpret our 
continuum-subtracted images.  

\begin{figure*}
\figurenum{3}
\plotone{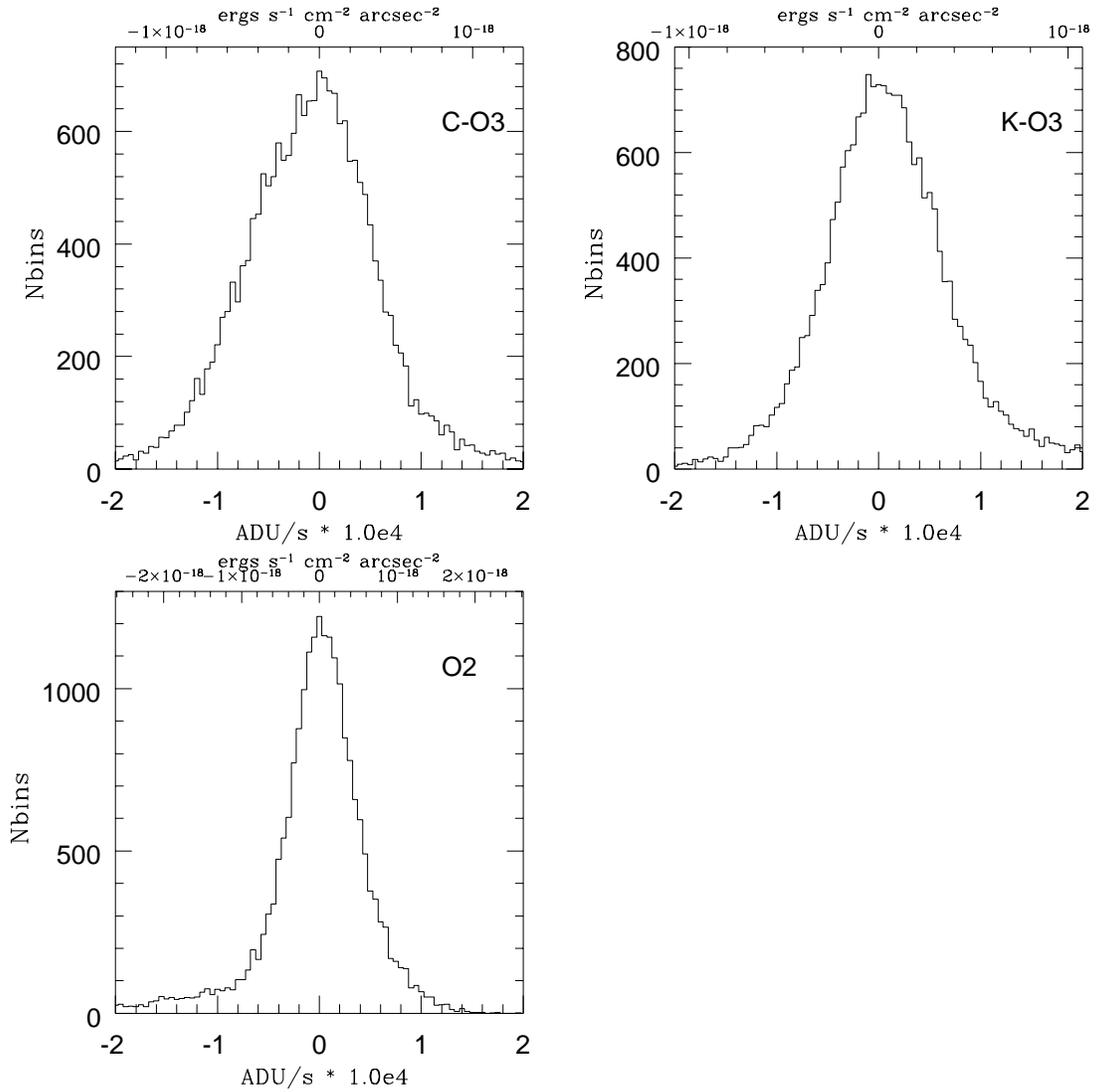}
\caption[Histogram of Sky Values]
{\label{fig:hist} 
The histogram of sky values for each of our block-averaged images, binned  
into intervals of 5.0 $\times 10^{-6}$ ADU/s~bin$^{-1}$.  Ideally, the sky values should all 
equal zero.  However, due to large-scale flat-fielding errors and the faint unmasked wings 
of stars and galaxies, there is usually a dispersion about zero in observed data.  This 
distribution determines the limiting surface brightness of our analysis. 
}
\end{figure*}

With the large-scale PSF determined, we re-masked all of the point sources in the 
frame, again down to a level of $0.1/t_{\rm exp}$~ADU\null.  We then masked out all of the
frame's non-point sources, using the SExtractor software package v2.2.2 \citep{SEXtractor} 
and the procedures of Feldmeier \etal~(2002, 2004).  Finally, after careful inspection,
we manually masked any remaining regions of flux (less than 1 percent of the remaining
area).  An illustration of the different steps of our masking procedure can be found 
in Figure~9 of \citet{feldmeier2002}.   

After applying this mask, we block-averaged the remaining pixels 
into $49 \times 49$ pixel ($13\farcs1 \times 13\farcs1$) 
superbins and calculated the robust mean of each superbin \citep{morrison1994}, 
while excluding from the analysis all superbins containing less than 
100 unmasked pixels.  This latter condition is rarely important, as it is only
used when a superbin contains an extremely bright star or a nearby galaxy.
In our blank-field images, the median number of pixels within each
superbin was 2125, 2010, and 1957 for the C-03, K-03, and O2 images,
respectively.  Given the large size of these superbins, each of their mean values 
should be equal to the sky subtracted mean of the image.  However, as the 
histograms of Figure~\ref{fig:hist} illustrate, there is a significant amount of 
dispersion about the sky subtracted mean.  This is the effect of large-scale 
flat-fielding errors, and the wings of stars and galaxies that remain unmasked 
despite our procedures.  

\begin{deluxetable*}{lccccccc}
\tablewidth{0pt}
\tablenum{3}
\tabletypesize{\scriptsize}
\tablecaption{Large scale Flat-Field Uncertainties\label{table:lff}}
\tablehead{
\colhead{Field}
& \colhead{Histogram FWHM\tablenotemark{a}}
& \colhead{Median Sky Value}
& \colhead{Uncertainty\tablenotemark{b}}
& \colhead{Exposure}
& \colhead{$N$\tablenotemark{c}}
& \colhead{Expected Poisson Error}
& \colhead{Flux error\tablenotemark{d}}\\
\colhead {} & \colhead{($10^{-4}$ADU s$^{-1}$ bin$^{-1}$)} & \colhead{(ADU s$^{-1}$ pixel$^{-1}$)} & \colhead {(percent)} 
& \colhead {Time (s)} & \colhead{}
& \colhead {($10^{-4}$ADU s$^{-1}$ bin$^{-1}$)} & \colhead {} 
} 
\startdata
C-O3  & 1.30 & 0.078 & 0.167 & 3600 & 24 & 0.122 & 8.7\\
K-O3  & 1.15 & 0.174 & 0.066 & 1200 & 47 & 0.224 & 6.2\\
O2    & 0.75 & 0.061 & 0.123 & 3600 & 36 & 0.088 & 9.9\\
\enddata
\tablenotetext{a}{See Figure~\ref{fig:hist} and the discussion in the text.}
\tablenotetext{b}{Defined as Histogram FWHM / Median Sky Value.}
\tablenotetext{c}{Effective number of exposures for each field.}
\tablenotetext{d}{final units are 10$^{-19}$ ergs s$^{-1}$ cm$^{-2}$ arcsec$^{-2}$}
\end{deluxetable*}

We conservatively define the large-scale flat-fielding error as the full-width at half-maximum 
of these distributions; 
these values are given in column 2 of Table~3.   As a cross-check of our
analysis, we also determined the standard deviation of the sky background at the edges of 
our LAE and point-source image stacks.  These values will not be identical to those derived 
from the superbins for two reasons: 1) the effect of scaling each individual image to a common flux level, 
and 2) the differing spatial distribution of the LAEs compared to the sky histograms.  Nevertheless, 
the values obtained from the sky background of the stacked images are within 50\% of the 
adopted values of Table~3.

The third and fourth columns of Table~3 compare the large-scale 
flat-fielding errors to the median sky values of the parent image.  These uncertainties
are comparable to those found for other deep imaging surveys \citep[\eg][]{feldmeier2004,
rudick2010}, but are not as good as the those associated with the original MUSYC 
broad-band survey \citep[see Table~5 of][]{gawiser2006a}.  We attribute this difference
to the lower flux values associated with our narrow-band data:  although the MUSYC
images were taken with the same telescope and reduced using similar techniques, the
increased count rate of the background sky depressed the relative importance
of additive sky terms.  By using the exposure times of each dataset (given in the fifth column), 
the effective number of exposures (given in the sixth column), and the
typical Mosaic~II CCD gain of 2.55 e$^{-1}$ per ADU, a representative
Poissonian error measured in ADU can be calculated for each superbin as follows:  

\begin{equation}
\sqrt{\frac{S}{T g A N}}
\end{equation}
where $S$ is the median sky value given in units of ADU~s$^{-1}$ per pixel, $T$ is the exposure time
in seconds, $g$ is the adopted gain of the CCDs, $A$ is the area of each superbin in pixels, 
(we adopted $49 \times 49$ pixels in this case) and $N$ is the effective number of exposures.
These uncertainties are given in the seventh column of Table~3.  
Comparing the two uncertainties, we
find that large-scale flat-fielding error is five to ten times larger than the
Poisson value.  This noise floor cannot be reduced by
simply including more sources in the stack, or binning the overall results: 
it is intrinsic to the data.  The final column in Table~3 gives 
this fundamental limit in units of monochromatic surface brightness. 

\section{Detailed Stacking}

With the systematic uncertainties of our datasets better determined, we began
a more detailed analysis of the surface brightness profiles of LAEs.  To do this, we
azimuthally averaged each stacked narrow-band image using a harsh ($2.5\sigma$) 
sigma-clipping algorithm, thereby creating a series of radial bins extending 
$18\farcs 7$ (70 pixels) from the source center.  Uncertainties for each radial bin were
calculated using a full error model, including readout noise, photon noise, small-scale flat-fielding
error, and large-scale flat-fielding error, and were determined in a similar manner as
in Feldmeier \etal~(2004; see their Appendix for a full description).  Next, we determined a new 
local sky value for each stack, by computing
the median pixel value in a circular annulus between  $13\farcs4$ and $18\farcs7$ (50--70 pixels)
from the LAE centroid. 
At $z \sim 2.1$, this annulus corresponds to a linear distance of between 111 and 156~kpc; 
at $z \sim 3.1$, the range is between 102 and 142~kpc.  In all cases, the local sky was 
within $\sim 1\%$ of the original sky value, and so the additional uncertainty introduced at this
step was small (on average a factor of two less than the large-scale flat-fielding error).
This constant sky value was then subtracted from the radial profiles.

\begin{figure*}
\figurenum{4}
\plotone{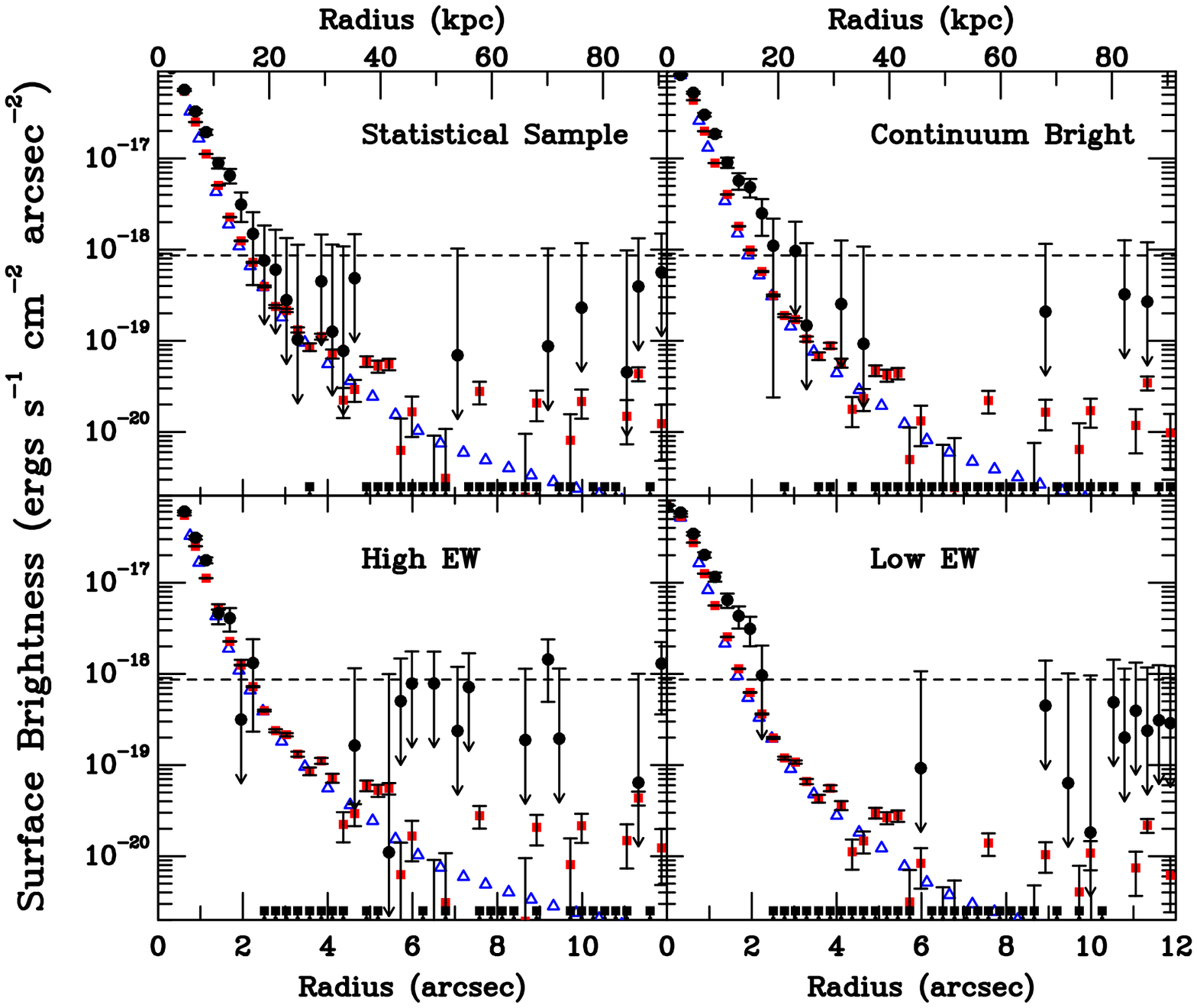}
\caption[C-O3 Radial Profiles]{\label{fig:co3prof} 
The radial profiles of the LAEs derived from the C-O3 sample.  For each of
these figures, the black circles represent the radial profile of the stacked LAEs, 
the blue open triangles represent the large-scale PSF, 
as determined in \S 4, and the red squares show the stacked radial profile of the stellar sources 
found in each frame.  LAE bins that have negative flux are denoted by a filled point at
the bottom of each plot.  The large-scale PSF and stellar sources have been normalized 
to match the three innermost points of the LAE profile.  The horizontal dashed line represents 
the flux limit comparable to our systematic uncertainties.  The differing profiles 
relate to the statistically complete sub-sample (top left), the UV Bright sub-sample (top right), 
the high EW sub-sample (bottom left) and the low EW sub-sample (bottom right), respectively.
}
\end{figure*}

\begin{figure*}
\figurenum{5}
\plotone{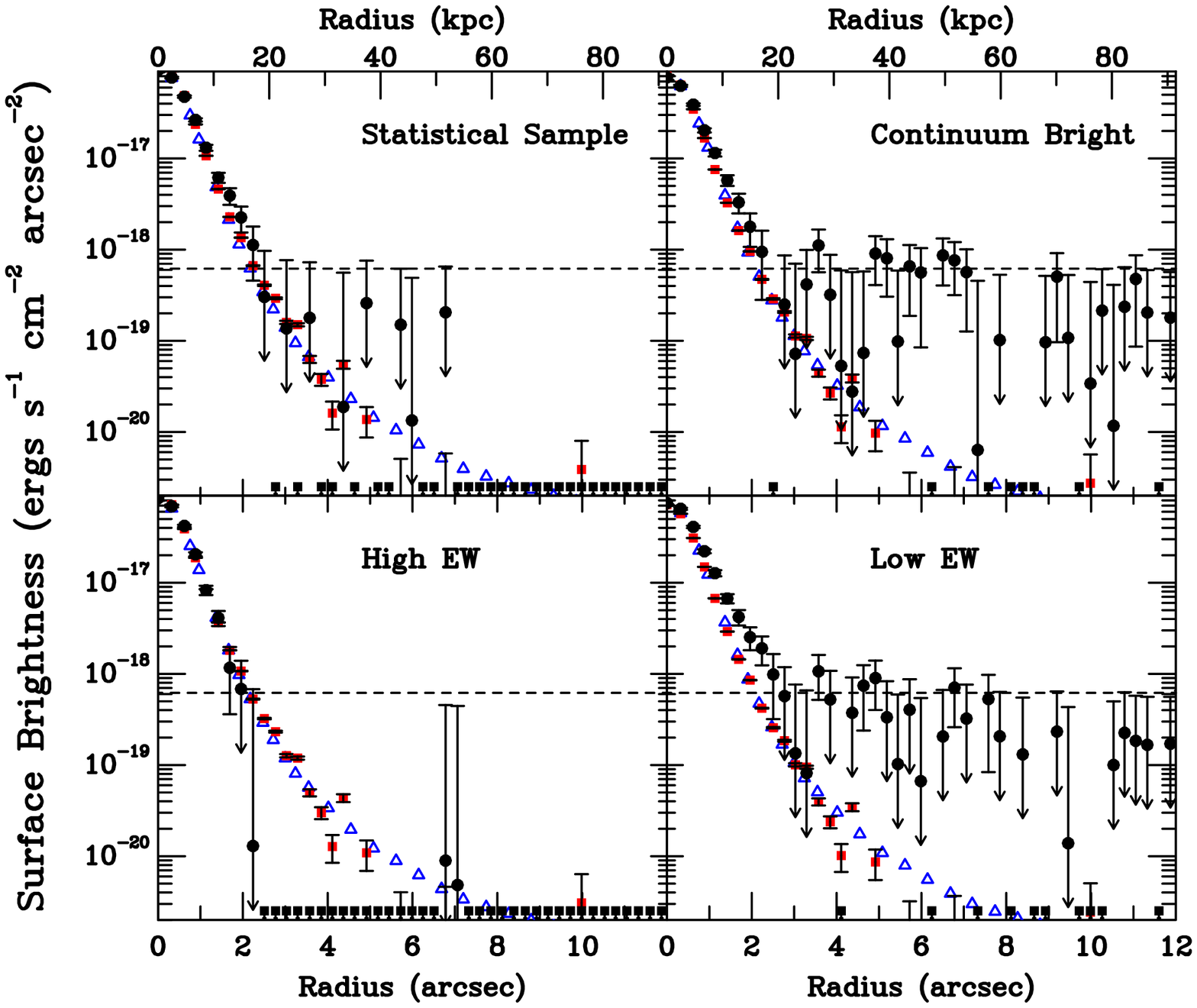}
\caption[K-O3 Radial Profiles]{\label{fig:ko3prof} 
The radial profiles of the LAEs derived from the K-O3 sample, similar to 
Figure~\ref{fig:co3prof}.}
\end{figure*}

\begin{figure*}
\figurenum{6}
\plotone{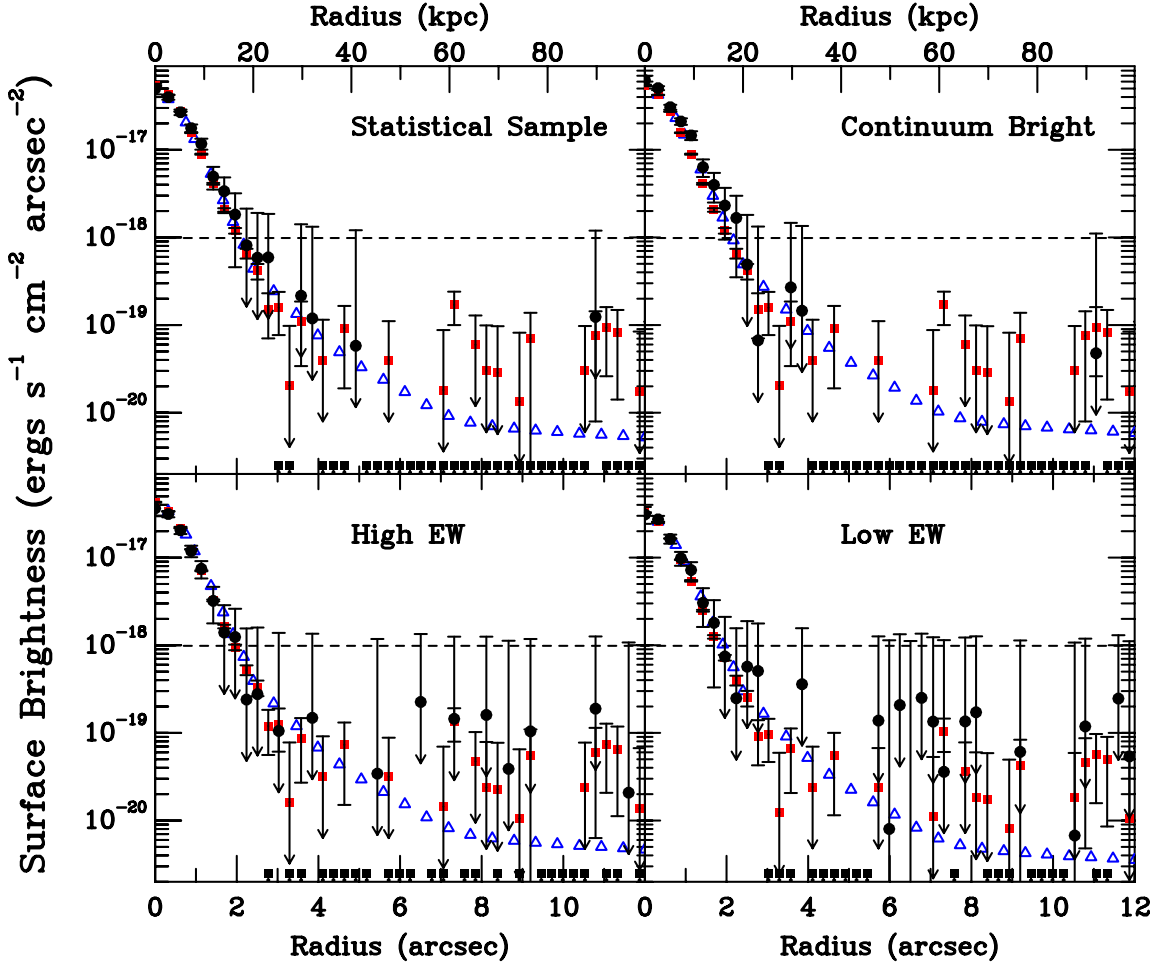}
\caption[O2 Radial Profiles]{\label{fig:o2prof} 
The radial profiles of the LAEs derived from the O2 sample, similar to 
Figure~\ref{fig:co3prof}.}
\end{figure*}

The resulting sky-subtracted radial profiles are plotted in Figures~\ref{fig:co3prof} 
through \ref{fig:o2prof}.  
Also shown in these figures are the frames' total PSFs (core and large-scale; calculated in \S 4), 
the radial profiles derived from the stacked set of 
point sources, and the surface brightness limits defined by the frames' large-scale 
flat-fielding errors.  In each figure, the comparison profiles have been scaled in flux to
match the three innermost points ($0\arcsec$--$0\farcs6$) of the corresponding LAE\null.
We note that the lines representing the surface brightness limits are approximate, since they
represent the mean flat-fielding error of the field.  Since LAEs exhibit angular clustering
\citep[\eg][]{francke2009}, the large-scale flat-fielding error appropriate to the stacked images may
be slightly different.  Nevertheless, they do illustrate the level at which stacks 
made from the wide-field Mosaic imager become suspect.

To better summarize the profiles displayed in Figures~\ref{fig:co3prof} through \ref{fig:o2prof}, we subtracted 
the point source profile from each LAE sub-sample stack, again using the three innermost points for scaling.  
Next, following S11 and M12, we fit the residuals to the exponential profile:
\begin{equation}
S(r) = C_{n} \exp(-r/r_{n})
\end{equation}
where $C_{n}$ is the normalization factor, and $r_{n}$ is the exponential scale length, 
using a Levenberg-Marquardt non-linear least squares algorithm \citep{press1992}, which
took into account negative fluxes for sky-subtracted bins.  The results of these fits are given in   
Table~4.  The uncertainties given in the table are the internal errors from the 
fit themselves, and are likely to be underestimates to the true uncertainties.   
For practical reasons, our fits were restricted to the spatial scales between $0\farcs 75$ 
(where seeing effects dominate) to $8\farcs 0$, where large-scale flatfielding  errors begin to make all 
measurements consistent with zero.  Moving the inner radius outward to match the $1\farcs 0$
value used by S11 does not qualitatively change the results.

\begin{deluxetable*}{lcccccc}
\tablewidth{0pt}
\tablenum{4}
\tablecaption{Point Source Subtracted Exponential Fits\label{table:expfit}}
\tablehead{
&\multicolumn{2}{c}{C-O3} &\multicolumn{2}{c}{K-O3} &\multicolumn{2}{c}{O2} \\
\colhead{Sample} &\colhead{$C_n$\tablenotemark{a}} &\colhead{$r_n$ (kpc)}
&\colhead{$C_n$\tablenotemark{a}} &\colhead{$r_n$ (kpc)}
&\colhead{$C_n$\tablenotemark{a}} &\colhead{$r_n$ (kpc)} }
\startdata
Statistically Complete & $3.6 \pm 1.1$ & $5.7 \pm 1.0$ & $3.1 \pm 1.5$ & $4.4 \pm 1.0$ & $1.3 \pm 1.6$ & $5.4 \pm 3.7$ \\
UV Bright       & $3.8 \pm 1.0$ & $6.0 \pm 0.9$ & $1.3 \pm 0.6$ & $6.4 \pm 1.8$ & $1.5 \pm 1.5$ & $5.7 \pm 3.2$ \\
High Equivalent Width  & $1.5 \pm 1.2$ & $5.5 \pm 2.6$ & $1.3 \pm 6.8$ & $2.8 \pm 5.0$ & $0.6 \pm 0.4$ & $3.7 \pm 12$ \\
Low Equivalent Width   & $2.4 \pm 1.1$ & $5.8 \pm 1.5$ & $1.2 \pm 0.3$ & $8.4 \pm 1.5$ & $0.4 \pm 1.6$ & $5.5 \pm 11$ \\
\enddata
\tablenotetext{a}{Flux is in units of $10^{-17}$~ergs~cm$^{-2}$~s$^{-1}$~arcsec$^{-2}$}
\end{deluxetable*}

\begin{figure*}
\figurenum{7}
\plotone{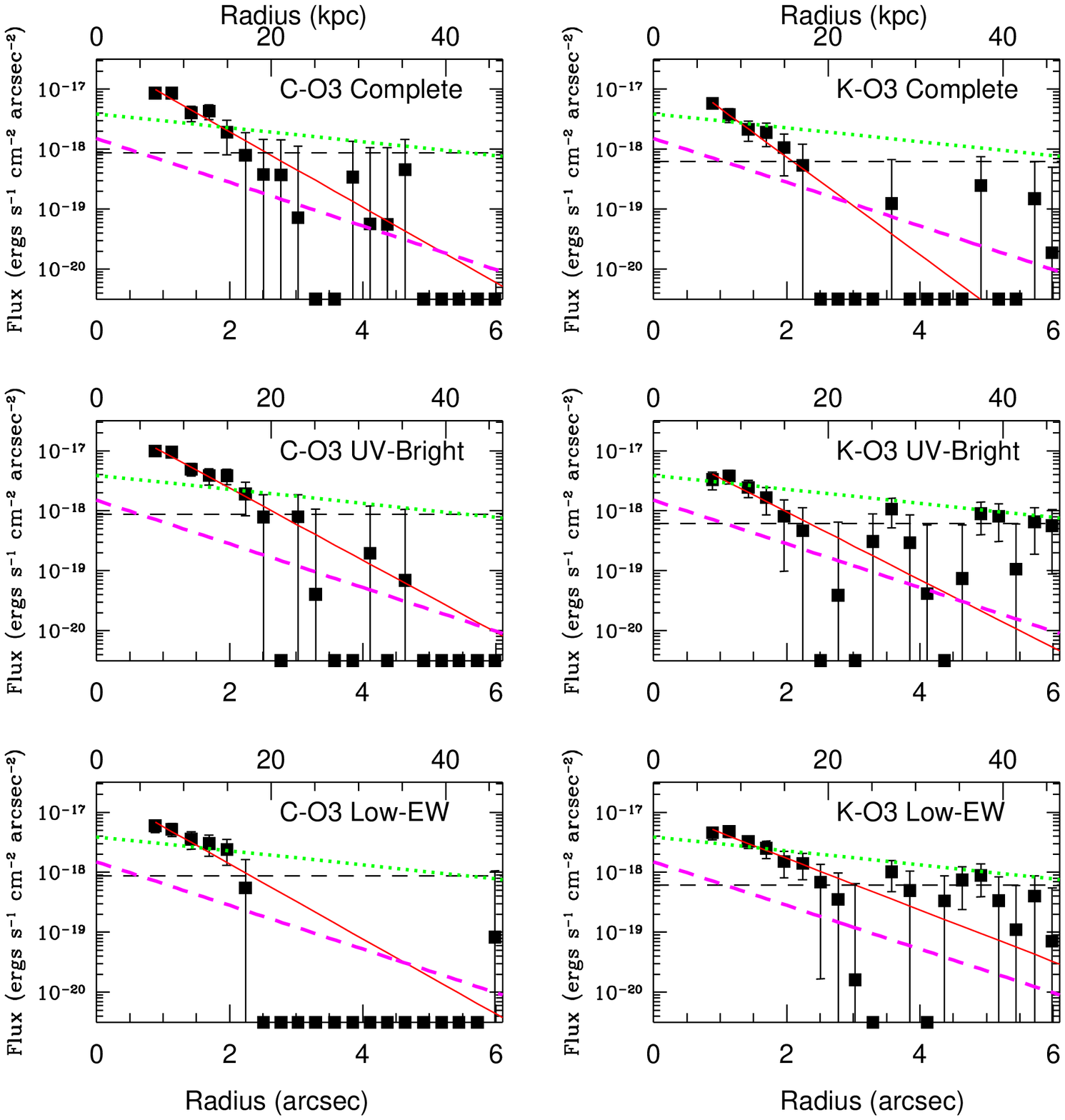}
\caption[Residuals]{\label{fig:residuals}
The residual radial profiles of our stacked LAEs, formed by subtracting 
a scaled version of each frame's point-source stack from the sub-sample's
original LAE stack.  Sub-samples which display no evidence for extended emission
are not shown.  Bins with negative flux are denoted by a filled point at the bottom of 
the plots. The best-fitting exponential model for each sub-sample (given in 
Table~4) is shown as the solid red line; for comparison purposes, 
the ``LAE only'' exponential fit of S11 (see their Table~2) is overplotted as the
green dotted line, and the ``$-1 < \delta_{LAE}^{a} < 0.5$'' (see their Table~1) exponential fit of M12 
is overplotted as the dashed magenta line.  Both of these fits are poor matches to the data.  
The horizontal dashed line in each profile represents the limit where systematic uncertainties
begin to dominate.}
\end{figure*}

An inspection of Figures~\ref{fig:co3prof} through \ref{fig:o2prof} and Table~4
reveals several interesting patterns.  At $z \sim 3$, there is evidence for the existence
of extended Ly$\alpha$ halos in some of the LAE sub-samples.  Specifically, the data suggest
that UV continuum bright LAEs are slightly more extended than the general LAE population,
and low-equivalent width objects have larger halos than their high-equivalent width counterparts.
Since the effect is present on stacks made from both the C-O3 and K-O3 frames,
this suggests that the effect is real, and that
there is a genuine difference between LAEs with strong and weak Ly$\alpha$ emission.
The sub-samples that show evidence for extended Ly$\alpha$ halos are plotted
in Figure~\ref{fig:residuals} in residual form, along with the best-fitting 
exponential profile.  For comparison purposes, the ``LAE only'' exponential fit of S11 and 
the ``$ -1 < \delta_{LAE}^{a} < 0.5$'' exponential fit of M12 are overplotted.  Our residuals
are a very poor match to these exponential fits: a simple $\chi^{2}$ test rejects both these fits 
at over 99\% confidence in all cases.  Our $\chi^{2}$ residuals to fits of the exponential profiles
found by S11 and M12 are given in Tables~5 and 6, respectively.  

\begin{deluxetable*}{lcccc}
\tablewidth{0pt}
\tablenum{5}
\tablecaption{Results of $\chi^{2}$ tests of residuals against S11 exponential fit\label{table:chisquare}}
\tablehead{
\colhead{Sample}
& \colhead{$\chi^{2}$}
& \colhead{Degrees of Freedom}
& \colhead{Reduced $\chi^{2}$}
& \colhead{Null Hypothesis Probability}
}
\startdata
C-O3 Complete  & 72.54 & 25 & 2.90 & 1.6 $\times$ 10$^{-6}$  \\
C-O3 UV Bright & 108.1 & 25 & 4.00 & 2.6 $\times$ 10$^{-12}$ \\
C-O3 Low-EW    & 56.24 & 25 & 2.55 & 1.3 $\times$ 10$^{-4}$ \\
K-O3 Complete  & 140.4 & 25 & 5.62 & 4.8 $\times$ 10$^{-18}$\\
K-03 UV Bright & 82.67 & 25 & 3.31 & 4.3 $\times$ 10$^{-8}$\\ 
K-O3 Low-EW    & 64.49 & 25 & 2.59 & 5.4 $\times$ 10$^{-5}$\\
\enddata
\end{deluxetable*}

\begin{deluxetable*}{lcccc}
\tablewidth{0pt}
\tablenum{6}
\tablecaption{Results of $\chi^{2}$ tests of residuals against M12 exponential fit\label{table:chisquare2}}
\tablehead{
\colhead{Sample}
& \colhead{$\chi^{2}$}
& \colhead{Degrees of Freedom}
& \colhead{Reduced $\chi^{2}$}
& \colhead{Null Hypothesis Probability}
}
\startdata
C-O3 Complete  & 94.54 & 25 & 3.78 & 5.1 $\times$ 10$^{-10}$  \\
C-O3 UV Bright & 134.9 & 25 & 5.40 & 4.8 $\times$ 10$^{-17}$ \\
C-O3 Low-EW    & 50.98 & 25 & 2.04 & 1.6 $\times$ 10$^{-3}$ \\
K-O3 Complete  & 51.02 & 25 & 2.04 & 1.6 $\times$ 10$^{-3}$\\
K-03 UV Bright & 49.72 & 25 & 1.99 & 2.3 $\times$ 10$^{-3}$\\ 
K-O3 Low-EW    & 74.51 & 25 & 2.98 & 8.1 $\times$ 10$^{-7}$\\
\enddata
\end{deluxetable*}

The reason for the poor agreement is not solely due to a mis-match in flux:
the best-fitting 
scale lengths for all our putative Ly$\alpha$ halos are relatively small, 
between $0\farcs 3$ and $1\farcs 0$ in angular units, 
or $\sim 3$ to $\sim 8$~kpc in physical size.   This is significantly less (a factor of
2-3) than the 20 to 25~kpc Ly$\alpha$ scale lengths found by S11 in their study of LBGs,
and only marginally consistent with the 9-–20 kpc scale lengths found in the LAE
samples studied by M12.  Formally, the derived scale lengths of all sub-samples are 
inconsistent with the S11 scale lengths at greater than 99\% confidence, 
and only the K-O3 UV bright and K-O3 low EW sub-samples are within 2$\sigma$ of the
M12 scale lengths.  This result is easily seen in Figure~\ref{fig:residuals},
where the best-fitting exponential profiles for our sub-samples are clearly steeper
than those of S11 and M12.  If the ECDF-S LAEs do have extended Ly$\alpha$
halos, they are much smaller than the halos found by other studies.

Finally, we note that $z \sim 2$ LAEs show no evidence for any LAE halos, as all
the profiles displayed in Figure~\ref{fig:o2prof} are all consistent with point-source
emission.  At face value, this is peculiar, since the effects of $(1 + z)^{4}$ cosmological surface 
brightness dimming should cause $z \sim 3$ halos to be more than $\sim 3$ times fainter than 
their $z \sim 2$ counterparts.  The slightly larger angular diameter at $z \sim 2$ also would 
tend to favor detection of extended emission.  While the poorer seeing and slightly brighter sky
of the O2 stacks may partly compensate for these effects, the lack of extended emission 
around the $z \sim 2$ objects may also indicate differences in the populations of
LAEs at $z \sim 2$ compared to $z \sim 3$.

\subsection{The effect of positional uncertainties}
Could some instrumental or sampling issue be affecting our conclusions?  Since
our analysis is differential in nature, and we are directly comparing the LAEs
to point sources using the exact same methods for stacking, averaging and fitting,
it seems unlikely that we are being misled by subtle systematic errors.  However,
one possible difference between the two samples lies in their mean observed flux.    
The LAEs measured in Figures~\ref{fig:co3prof} 
through \ref{fig:o2prof} are, on average, more than 100 times fainter than their point source
counterparts, and hence have larger positional uncertainties.  Since the act of stacking relies on
knowing the objects' centroids, it is possible that in our co-addition procedure, we
have artificially introduced a ``halo'' into our LAE sample.

To quantify the potential strength of this effect we performed a series of Monte-Carlo experiments, each time
adding 200 faint ($S/N = 9$) artificial point sources to the C-O3 frame, and stacking the
results, while adding a random amount of positional uncertainty, $\sigma_{\rm pixel}$, into the
assumed $x$ and $y$ coordinate of each object.  Our choice of $\sigma_{\rm pixel}$ spanned the entire 
range of possible centroiding errors (0.5, 0.75, 1.0, 1.5, and 2.0 pixels), while our flux assignments 
were chosen to isolate the effects of positioning from that due to the effects of low signal-to-noise.  
We stacked, averaged, sky subtracted, and fit these faint artificial point sources in the exact same 
way as before, and examined the behavior of $C_n$ with $\sigma_{\rm pixel}$.
Any non-zero values of $C_n$ would indicate an additional uncertainty above
that found from the photometric errors.    

\begin{figure}
\figurenum{8}
\plotone{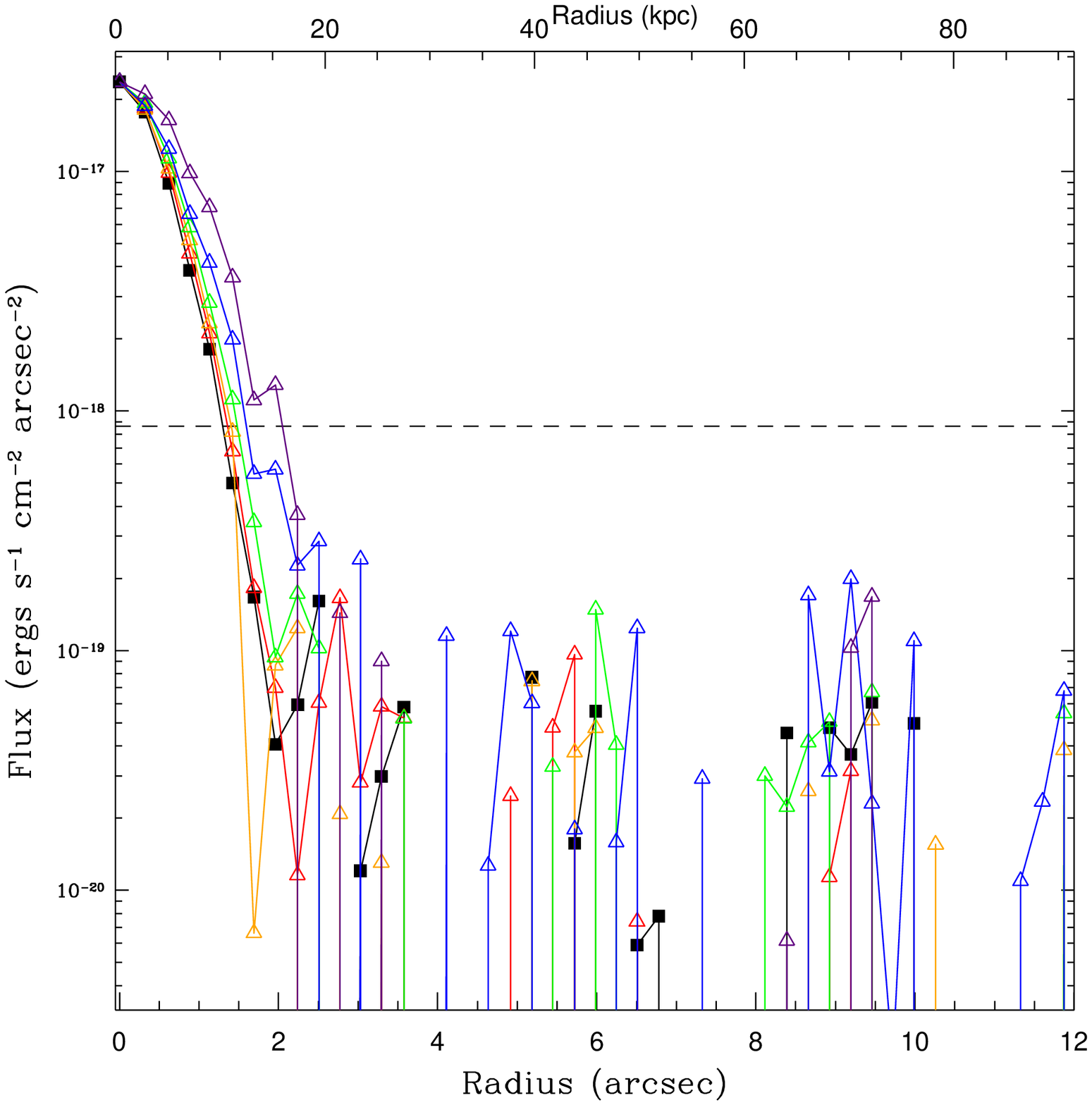}
\caption[Centroid errors]{\label{fig:centroid} 
The effects of centroiding errors on our measured profiles.  The black filled points
represent the co-added profile of 200 artificial point sources (each with a signal-to-noise of 9)
on the C-O3 image.  The open triangles represent the same data, but with differing 
amounts of $\sigma_{pixel}$, with red, orange, green, blue, and purple (running from closest
to furthest away from the solid squares)
corresponding to $\sigma_{pixel} = 0.5$, 0.75, 1.0, 1.5, and 2.0 pixels, respectively.
Each profile has been normalized to the central point for comparison purposes.}
\end{figure}

\begin{deluxetable}{lccc}
\tablewidth{0pt}
\tablenum{7}
\tabletypesize{\small}
\tablecaption{Effects of Centroiding Errors on Exponential Fits to Point Sources\label{table:expuncertain}}
\tablehead{
\colhead{$\sigma_{pixel}$}
& \colhead{$C_{n}$\tablenotemark{a}}
& \colhead{$r_{n}$}
& \colhead{$r_{n}$\tablenotemark{a}}\\
\colhead {(pixels)} & \colhead {(10$^{-17}$ ergs s$^{-1}$ cm$^{-2}$ arcsec$^{-2}$)} & \colhead{(arcsec)}
& \colhead{(kpc)}} 
\startdata
0.0                         & -0.05 $\pm$ 0.32 & 1.3 $\pm$ 6.2 &  11 $\pm$ 52\\
0.5                         &  0.5  $\pm$  4.7 & 0.4 $\pm$ 1.7 & 3.3 $\pm$ 14 \\
0.75                        &  1.1  $\pm$  6.7 & 0.4 $\pm$ 0.9 & 3.3 $\pm$ 7.3\\
1.0                         &  1.0  $\pm$  3.6 & 0.5 $\pm$ 0.7 & 4.2 $\pm$ 5.8\\
1.5                         &  0.8  $\pm$  1.7 & 0.6 $\pm$ 0.6 & 5.0 $\pm$ 5.1\\  
2.0                         &  1.2  $\pm$  1.8 & 0.6 $\pm$ 0.4 & 5.0 $\pm$ 3.6\\
\enddata
\tablenotetext{a}{The meaning of these parameters are identical to Table~4}
\end{deluxetable}

As illustrated in Figure~\ref{fig:centroid} and summarized in Table~7,  
this positional broadening has the potential to mimic the effects of a diffuse halo: 
large ($\sim 1$~pixel) uncertainties in the $(x,y)$ positions of individual LAEs can 
cause a point-source stacked image to appear extended, and generate a false exponential profile.  
Obviously, the larger the value 
of $\sigma_{\rm pixel}$, the larger the size of the apparent halo, so the amplitude of this 
effect depends strongly on the brightness distribution of the LAE sample being studied.

To determine this quantity, we adopted the LAE flux distribution given in \citet{gronwall2007},
and measured the uncertainty in position as a function of signal-to-noise using
a series of artificial star experiments.  Our analysis showed that the median coordinate 
uncertainty for objects near the bright-end of the LAE luminosity function is very small, 
$\sim 0.05$ pixels, while that for objects at the faint end of the statistically complete 
sample is $\sim 0.5$~pixels in the median, and, at worst, $\sim 0.8$~pixels.  
The curves of 
Figure~\ref{fig:centroid} then suggest that errors in the LAE centroiding can explain
some ($\sim 25\%$ of the flux), but not all, of the excesses we detect.

Finally, it is unlikely that the discrepancy between our results and those of S11 and M12 comes from 
the small differences in the stacking procedures adopted by each study.  Our median combination
and radial sigma-clipping algorithm has the advantage of being robust against unwanted flux from nearby 
continuum sources, and is less prone to contamination by unusual objects 
(such as Lyman-alpha blobs) than other methods.  Yet the consistent scale lengths found 
by S11 across all their sub-samples make it unlikely that their result is driven by 
contamination, and, in any event, their masking procedure should have effectively 
removed any outlying flux.  Similarly, the median stacks computed by M12 should have 
also been immune against contamination by interlopers.  Their method of combining 
sources without any scaling or object weighting is slightly sub-optimal in terms of
signal-to-noise, but the much larger size of their LAE sample should have
overcome this limitation.  

We therefore conclude that we have some, but not conclusive, evidence for Ly$\alpha$
halos in our sub-samples.  At $z \sim 3$, LAEs with low equivalent widths and bright 
continuum  magnitudes display more evidence for the presence of extended Ly$\alpha$
emission, but due to the uncertainties in flatfielding, sky subtraction, and coordinate 
centroiding, the detection cannot be considered definitive.  Nevertheless, we can say that if 
Ly$\alpha$ halos do exist in our samples, they are much smaller in physical size 
than those found by S11 and M12\null.  Moreover, there are clear differences between our 
results at $z \sim 3$ and $z \sim 2$.  Despite the reduced effects of cosmological surface 
brightness dimming, we see no evidence for any extended halo at the lower redshift.

\section{Determining Halo Detection Limits Through Monte-Carlo Analysis}

Figure~\ref{fig:hist} and Table~3 illustrate the fundamental limitation
of all LAE surface photometry performed over scales greater than
$\sim 100$~kpc.  It does not address the question of small-scale background fluctuations,
and it is at least conceivable that on galaxy-sized scales, the uniformity of the sky background
improves.  To test this possibility, we performed a series of Monte Carlo simulations, using
model LAE halos of the form suggested by S11.  The goal of these simulations was to ensure that
our imaging has sufficient signal-to-noise to detect extended LAE halos, if they are present.

We began by using the GALFIT analysis package \citep{galfit2002, galfit2010}
to create a set of artificial LAEs, with each LAE consisting of two components:
a point source core, and an exponential halo of the form suggested by S11 (with an adopted scale 
length, $r_{n} = 28.4$~kpc). In each simulation, we chose a global 
halo-to-core flux ratio (corresponding to differing values of $C_{n}$), convolved our composite
galaxies with the frame's large-scale PSF,  
and assigned these simulated LAEs random fluxes drawn from the \citet{schechter1976} 
luminosity function appropriate for their redshift
\citep{ciardullo2012}.  We then randomly placed the artificial galaxies on their respective frames
and stacked those objects brighter than the frame's 90\% completeness limit.  We performed
eleven such simulations for the C-O3 and the O2 images, with halo-to-core ratios spaced
evenly between zero and four.

\begin{figure*}
\figurenum{9}
\plotone{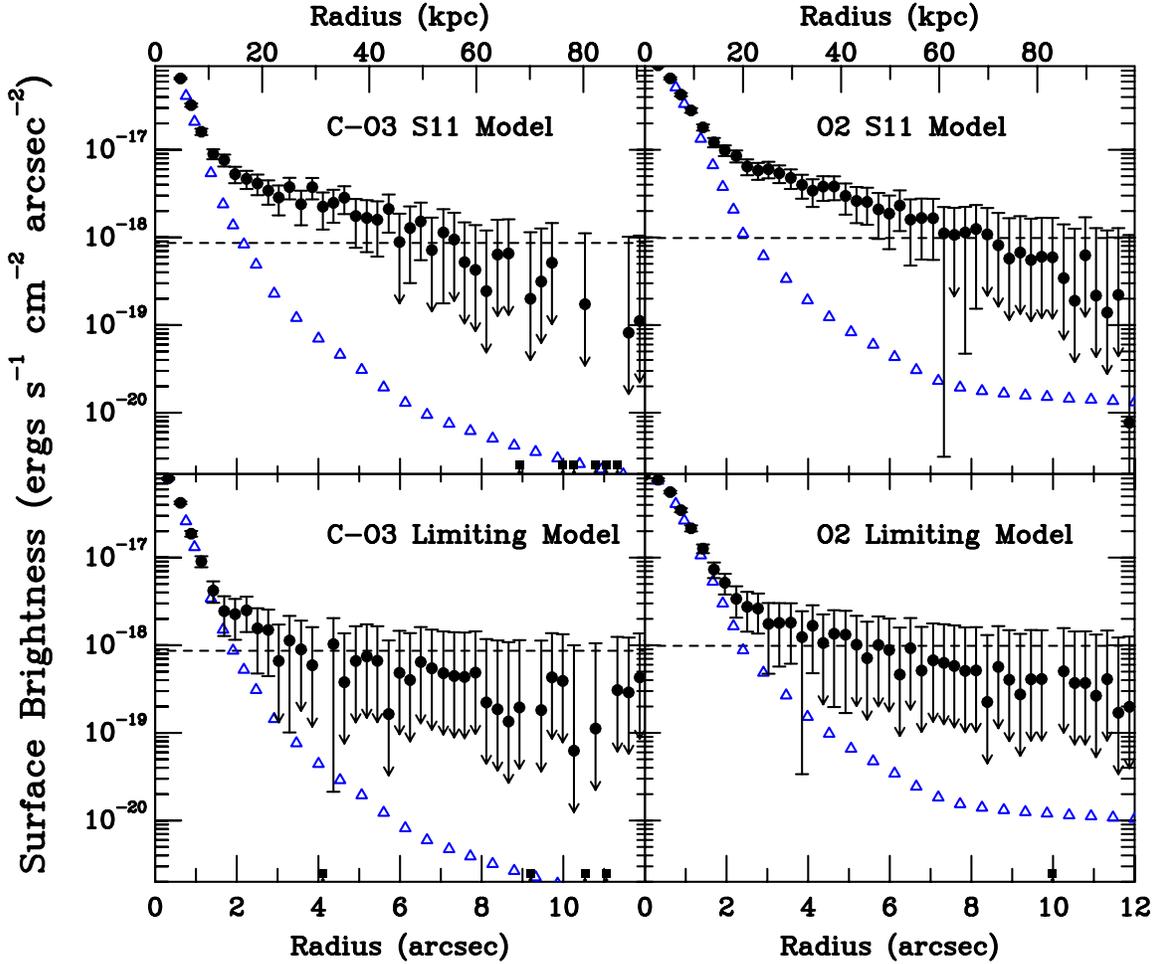}
\caption[Monte Carlo CO3]{\label{fig:montecarlo1}
A comparison of Monte-Carlo models with an exponential halo added to the large-scale PSF.  
The meaning of the symbols is identical to Figure~\ref{fig:co3prof}.  The top row indicates 
the halo-to-core ratio consistent with the results of S11 ($r_{n}$ = 28.4 kpc).  An extended halo of this 
type would have been clearly detected in our data.  The bottom row indicates the models
where an extended halo is just barely detected, with a halo-to-core ratio of $\approx 1.45$ in the C-O3 plot, 
and $\approx 1.09$ in the O2 plot. }
\end{figure*}

The top panels of Figure~\ref{fig:montecarlo1} show the results of these simulations
for halos of the type suggested by S11.  It is clear that at this halo-to-core ratio (corresponding
to $C_{n} \approx 4.6 \times 10^{-18}$~ergs~cm$^{-2}$~s$^{-1}$~arcsec$^{-2}$  for the C-O3 stacks,
and $C_{n} \approx 1.1 \times 10^{-17}$~ergs~cm$^{-2}$~s$^{-1}$~arcsec$^{-2}$ for the O2 stacks), 
extended Ly$\alpha$ emission would have been easily detected, and their
morphology would have been quite different from the observed profiles displayed in 
Figures~\ref{fig:co3prof}-\ref{fig:residuals}.  The panels also demonstrate that the
surface brightness limits to which we can confidently trace Ly$\alpha$ emission are similar
to those predicted by the large-scale flat-fielding error.  Even at scales
less than $\sim 100$~kpc, the systematic effects discussed in \S4 are still the dominant error 
terms in our analysis.

Of course, it is certainly possible that the Ly$\alpha$ halos of high-redshift LAEs are
fainter than those described by S11, and have correspondingly smaller values of $C_n$. 
To examine this possibility, we visually inspected and repeated our non-linear
least squares analysis on simulated halos with smaller halo-to-core ratios.  The bottom 
panels of Figure~\ref{fig:montecarlo1} illustrate those models with marginally-detected halos.  
In the C-O3 panel, we found a limiting value of 
$C_{n} \approx 2 \times 10^{-18}$~ergs~cm$^{-2}$~s$^{-1}$~arcsec$^{-2}$, 
while in the O2 plot, the limiting value of 
$C_{n}$ is  $\approx 4 \times 10^{-18}$~ergs~cm$^{-2}$~s$^{-1}$~arcsec$^{-2}$
Any LAE halos with fluxes below these limits would be undetectable on our images.  
We note that the stacked Ly$\alpha$ halo found by S11 for LBGs is relatively bright, 
as it contains $\sim 3.4$ times more flux than the galaxy's central core.  The halos detected 
by M12, however, are substantially fainter, with best fit values for $C_n$ 
ranging from $3.2$ down to $0.7 \times 10^{-18}$~ergs~cm$^{-2}$~s$^{-1}$~arcsec$^{-2}$.  
Therefore, in our LAE samples, the relatively bright halos suggested by S11 are ruled out, 
while the fainter halos claimed by M12 may still exist.

\section{Discussion}
Most cosmological simulations -- even those which include radiative transfer from Ly$\alpha$ 
-- do not include the effects of resonant scattering of photons from the inter- and 
circum-galactic medium.   One exception is the work of \citet{zheng2011}, who calculated 
the expected spatial distribution of extended Ly$\alpha$ emission from LAEs at $z \sim 6$.    
These models predict that in the Ly$\alpha$ emission line, a high-redshift LAE should have a 
central cusp, and a slowly declining envelope which extends outward from $\sim 0.2$~Mpc 
to $\sim 1$~Mpc (in co-moving distance).    
Unfortunately, it is difficult to directly compare our observational results at $2 < z < 3$ to
the predictions of these models.  The distribution of gas at $z \sim 6$ is largely defined
by infall, and is related to the virial radius of the galaxy; both these parameters evolve
rapidly in the Gyr between $z \sim 6$ and $z \sim 3$.  In addition,   
the \citet{zheng2011} models have a poorer spatial resolution than our data, making it difficult to compare except at
the largest radii.  These simulations also do not include the effects of dust, and, depending 
on how this component is 
distributed, its effect on the propagation of Ly$\alpha$ photons can be dramatic 
\citep[\eg][]{neufeld1991, fink2009}.  Finally, the \citet{zheng2011} models do not
take into account the internal structure and kinematics of the ISM\null.  Thus, it is not possible 
to directly compare their model predictions with observational results.

More recently, \citet{dijkstra2012} performed radiative transfer calculations for Ly$\alpha$ 
photons propagating through clumpy, dusty, large scale outflows.  While these authors
do predict the existence of extended Ly$\alpha$ halos, their initial models (I--III) have halos
that are more concentrated (inner scale lengths of $\sim$ 4 kpc) than those 
observed by S11, and are fainter by almost an order of magnitude 
($\approx 4.5 \times 10^{-20}$~ergs~cm$^{-2}$~s$^{-1}$~arcsec$^{-2}$ at a radius of $\approx$ 5\arcsec).  
More complex models (IV--V) that include the deceleration of clumps are better matches to the halos
observed by S11, but still underpredict the flux by a factor of 2--3.  
Unfortunately, due to our observational limits, we would not be able to detect 
halos of these types; only the two innermost points of their halos would be measurable in our data, with 
fluxes of $0.1-1 \times 10^{-17}$~ergs~cm$^{-2}$~s$^{-1}$~arcsec$^{-2}$.  
Several recent studies of Ly$\alpha$ emitters within a cosmological framework explicitly 
include explicit radiative transfer, and these could be applied to the problem of Ly$\alpha$ 
halos \citep{forero2011,garel2012,orsi2012}. 

As detailed above, accurate surface photometry at low flux levels is difficult, and fraught 
with systematic errors.   Consequently, direct comparisons with other
Ly$\alpha$ halo measurements is difficult and may ignore important systematic effects.   
Using long-slit spectroscopy, \citet{rauch2008} 
found extended Ly$\alpha$ emission in roughly half of a sample of 27 LAEs 
in the redshift range $2.7 \leq z \leq 3.8$.  But this emission was weak: no brighter than 
$\sim 1.8 \times 10^{-19}$ erg~s$^{-1}$~cm$^{-2}$~arcsec$^{-2}$ at $\sim 4\arcsec$
($\sim 30$~kpc) and well below the level where systematic uncertainties dominate our
error budget.  Yet a similar study by \citet{cassata2011} on 217 LAEs in the redshift range 
$2 \leq z \leq 6.6$ found no evidence for extended Ly$\alpha$, as all their objects 
were found to be compact (FWHM $\sim 0\farcs 9$).  M12, by using the stacked images of 
several hundred field LAEs found on Subaru frames, found halos at the level of 
$1 \times 10^{-18}$~ergs~s$^{-1}$~cm$^{-2}$~arcsec$^{-2}$.  For comparison, the LBG 
halos of S11 and \citet{hayashino2004}, which were measured in highly overdense regions of the 
universe, are almost an order of magnitude brighter.  Other attempts to image LAE 
halos, both from the ground \citep{smith2012}, and from space \citep{bond2010, fink2011c}
have placed only weak upper limits on the phenomenon, and generally probe very different 
spatial scales.  Recently, \citet{jiang2013} has stacked LAEs at z $\simeq$ 5.7 and 6.5  
and has not found signs of extended LAE halos.

S11 hypothesize that extended Ly$\alpha$ halos are a ``generic property of high redshift 
star-forming galaxies'', while M12 argue that the emission properties of Ly$\alpha$ 
halos depend on environment.  Neither of these studies took into account the systematics
associated with the large radius point-spread-function or large-scale flat fielding, but
neither are in direct conflict with our results.   The median LAE in our analysis is over a 
magnitude fainter than the LBGs studied by S11, and our surface brightness limits, 
$9.9 \times 10^{-19}$ erg~s$^{-1}$ cm$^{-2}$ arcsec$^{-2}$  for $z\sim 2.1$ and 
$8.7 \times 10^{-19}$ and $6.2 \times 10^{-19}$ erg~s$^{-1}$ cm$^{-2}$ arcsec$^{-2}$  
for the two $z\sim 3.1$ samples are brighter than the values quoted by M12.  Thus, 
there are several ways to reconcile the three surveys.

First, LAEs may have significantly fainter Ly$\alpha$ halos than Lyman-break galaxies.
In general, the star formation rates of LAEs are an order of magnitude smaller than
those of LBGs \citep[\eg][]{gronwall2007}, and it is reasonable to hypothesize that the lower 
masses associated with LAEs also imply less circum-galactic material.  These two factors may
combine to produce Ly$\alpha$ halos whose surface brightness is highly sensitive to the
luminosity of the underlying galaxy.   This argument is somewhat mitigated by the fact that
Lyman-break galaxies have more internal extinction than their LAE counterparts, but
there are ways to distribute this dust so that the resonantly scattered Ly$\alpha$ photons
can escape their surroundings \citep{neufeld1991, fink2009}.   The fact that our best 
evidence for extended emission occurs on the C-O3 and K-O3 stacks of LBG-like objects 
and low-equivalent width sources argues in support of this interpretation.

Alternatively, the observability of Ly$\alpha$ halos may be due largely to galactic
geometry.   The probability of detecting Ly$\alpha$ emission from a galaxy depends on
many factors, including viewing angle, dust content, gas dynamics, and intrinsic
geometry.  By definition, LAEs are systems where the Ly$\alpha$ escape path falls 
along our line-of-sight; in galaxies selected via the Lyman-break technique, Ly$\alpha$
may leak out in other directions, thus facilitating the detection of a ``halo.''  S11's data
argue against this interpretation, since the halos surrounding LBGs with Ly$\alpha$
emission are no different from those where Ly$\alpha$ was only seen in absorption. 
However, the Ly$\alpha$ equivalent widths of typical LAEs are much greater than those
associated with LBGs, so this selection effect may still play a role.  Recently,
\citet{verhamme2012} have claimed that Ly$\alpha$ emission from galaxies can have strong 
inclination effects, with LAEs with high equivalent widths being more 
likely to be found in face-on, rather than edge-on orientations.  Although there are many details
still left to be resolved, it is also worth noting that the simulations of \citet{verhamme2012}
show a substantially weaker halo than S11.

A third possibility concerns environment.  M12 have argued that a galaxy's surroundings play an important
role in the observability of Ly$\alpha$ emission.   The LBGs studied by S11 are primarily located in overdense regions of the 
universe, including the well-known field SSA22 \citep{steidel1998} and two regions centered 
on hyper-luminous QSOs (HS1700+64 and HS1549+195).   In these ``proto-cluster''
environments, there is likely to be a large amount of circum-galactic material, which could
scatter the Ly$\alpha$ photons which escape from galaxies.  Indeed, not only is 
SSA22 known for possessing large numbers of luminous Ly$\alpha$ blobs 
\citep[e.g.,][]{matsuda2004, steidel2000}, but, according to M12, its LAEs are similar in
strength and profile to those found by S11 for the region's LBGs.  In lower
density regions, M12 found the Ly$\alpha$ halos of LAEs to have steeper profiles 
with less total luminosity.  
The LAEs studied in this paper are all in the blank-field environment of the ECDF-S, which is 
equivalent to the $ -1 < \delta_{LAE}^{a} < 0.5$ sub-sample of M12.
If this is true, then the lack of halo emission around our galaxies should not be too 
surprising.

Finally, we note that our Ly$\alpha$ halo measurements at $z \sim 3.1$ differ from those at
$z \sim 2.1$:  despite the $(1+z)^4$ gain in surface brightness,  we see no evidence for 
extended Ly$\alpha$ halos in any of our lower-redshift sub-samples.  One possible 
explanation for this difference lies in the uniformity of LAE populations.   A number of 
analyses have shown that LAEs at $z \sim 2.1$ are, on average, slightly larger, dustier, 
and more heterogeneous than their $z \sim 3.1$ counterparts \citep{nilsson2009,nilsson2011,guaita2011,bond2012}.  
It is possible that this diversity at lower redshift might also lead to differences 
in halo properties.

\section{Conclusion}

Surface photometry at very low flux levels is treacherous, and extreme care must be
used in modeling systematics such as the large-radius point-spread-function and the
large-scale flat-fielding errors.  By taking such factors into account, we were able to
form stacked profiles of three sets of LAEs in the Ly$\alpha$ emission-line and
to obtain robust estimates of their uncertainties.  At $z \sim 3.1$, we 
find mixed evidence for the presence of extended Ly$\alpha$ halos down to a 
surface brightness limit of $\approx 7 \times 10^{-19}$~ergs~cm$^{-2}$~s$^{-1}$~arcsec$^{-2}$.
At $z \sim 2.1$, where cosmological dimming is a factor of 3.2 less, we see no extended
emission down to surface brightnesses of $\approx 1 \times 10^{-18}$~ergs~cm$^{-2}$~s$^{-1}$~arcsec$^{-2}$.   
This is the first time the rigorous techniques of low-flux level surface photometry 
have been applied to the problem of galactic emission-line halos.

While our observations are not in direct conflict with those of \citet{steidel2011} and 
\citet{matsuda2012}, they do place important constraints on the systematic behavior of 
Ly$\alpha$ halos and imply that such studies have higher uncertainties than originally thought.  
However, further progress on this problem will be difficult in the near-term.       
To succesfully image the emission-line halos around LAE and LBG halos, one should obtain 
deep narrow-band exposures on telescopes with wide-field imagers that also have low 
scattered light profiles and very stable flat-fields.   This means paying close attention to 
the details of baffling and flat-fielding (i.e., taking the dark-sky flat flat-fields at the 
same hour angle as the observations) and using closed-tube telescopes with cameras designed 
to minimize the total number of internal reflections.  Our observations, as well as those of 
S11 and M12, were performed with open-tube telescopes and wide-field imagers that 
contained multi-component optical corrector systems.  As this is the norm for large wide-field 
imaging telescopes, it will be difficult to extend imaging surveys much below the 
surface brightness limits achieved here, though observations of additional fields will allow 
for better determination of systematic effects and allow us to better probe the effects of
galactic environment on LAE halos.

Alternatively, it is possible to study the emission-line halos of high-redshift galaxies
with spectroscopy.  Starting in 2014, the integral-field units of the Hobby-Eberly 
Telescope Dark Energy Experiment (HETDEX) will obtain two-dimensional spectra for 
$\sim 10^6$ Ly$\alpha$ emitters in the redshift range $1.9 < z < 3.5$ \citep{hill2007}.
These data, which will have $\sim 0\farcs 5$ spatial resolution and $\sim 6$~\AA\ spectral
resolution, should allow us to trace circum-galactic Ly$\alpha$ emission to levels far
below what is currently possible and to determine the properties of these systems as
a function of luminosity, equivalent width, dust content, redshift, and environment, and
thus illuminate material in the high-redshift universe's cosmic web.

\acknowledgements

This work was supported by NSF grants AST 06-07416, AST 08-07570, AST 08-07873, 
AST 08-07885, and AST 10-55919, and DOE grants DE-GF02-08ER41560 and DE-FG02-08ER41561. 
The Institute for Gravitation and the Cosmos is supported by the Eberly College 
of Science and the Office of the Senior Vice President for Research at 
the Pennsylvania State University.

We acknowledge valuable discussions with Lennox Cowie, Esther Hu, Sangeeta Malhotra, 
James Rhoads, and Tomoki Saito.  We also thank the anonymous referee for several suggestions that
improved the quality of this paper.  We thank the staff of Cerro Tololo Inter-American 
Observatory for their assistance with our observations. This research has made use of NASA's 
Astrophysics Data System.

{\it Facility:} \facility{Blanco: (MOSAIC II)}


\clearpage

\begin{thebibliography}{}

\bibitem[Acquaviva \etal(2011)]{acquaviva2011} Acquaviva, V., Gawiser, E., \& Guaita, L. 
2011, \apj, 737, 47 

\bibitem[Adams et al.(2011)]{adams2011} Adams, J.~J., Blanc, 
G.~A., Hill, G.~J., et al.\ 2011, \apjs, 192, 5 

\bibitem[Adelberger \etal(2003)]{adelberger2003} Adelberger, K.L., 
Steidel, C.C., Shapley, A.E., \& Pettini, M. 2003, \apj, 584, 45 

\bibitem[Adelberger \etal(2005)]{adelberger2005} Adelberger, K.L., 
Steidel, C.C., Pettini, M., Shapley, A.E., Reddy, N.A. \& Erb, Dawn K. 2005, \apj, 619, 697 

\bibitem[Altmann \etal(2006)]{altmann2006} Altmann, M., M\'endez, R.A., 
van Altena, W., Korchargin, V., \& Ruiz, M.T. 2006, Rev.~Mex.~A.A., 26, 64 

\bibitem[Barger et al.(2012)]{barger2012} Barger, A.~J., Cowie, 
L.~L., \& Wold, I.~G.~B.\ 2012, \apj, 749, 106

\bibitem[Barnes et al.(2011)]{barnes2011} Barnes, L.~A., Haehnelt, 
M.~G., Tescari, E., \& Viel, M.\ 2011, \mnras, 416, 1723 

\bibitem[Bernstein(2007)]{bernstein2007} Bernstein, R.A. 2007, \apj, 666, 663 

\bibitem[Berry \etal(2012)]{berry2012} Berry, M., Gawiser, E., Guaita, L.,  Padilla, N., 
Treister, E., Blanc, G.A., Ciardullo, R., Francke, H., \& Gronwall, C. 2012, \apj, 749, 4

\bibitem[Bertin \& Arnouts(1996)]{SEXtractor} Bertin, E., \& Arnouts, S. 1996, 
\aaps, 117, 393 

\bibitem[Blanc et al.(2011)]{blanc2011} Blanc, G.~A., Adams, 
J.~J., Gebhardt, K., et al.\ 2011, \apj, 736, 31

\bibitem[Bond \etal(2010)]{bond2010} Bond, N.A., Feldmeier, J.J., Matkovi\'c, A., 
Gronwall, C., Ciardullo, R., \& Gawiser, E. 2010, \apjl, 716, L200 

\bibitem[Bond et al.(2011)]{bond2011} Bond, N.~A., Gawiser, E., 
\& Koekemoer, A.~M.\ 2011, \apj, 729, 48 

\bibitem[Bond \etal(2012)]{bond2012} Bond, N., Gawiser, E., Guaita, L., 
Padilla, N., Gronwall, C., Ciardullo, R., \& Lai, K. 2012, \apj, 753, 95

\bibitem[Cassata et al.(2011)]{cassata2011} Cassata, P., Le F{\`e}vre, O., Garilli, B., et al.\ 2011, \aap, 525, A143

\bibitem[Ciardullo \etal(2012)]{ciardullo2012} Ciardullo, R., Gronwall, C., Wolf, C., 
McCathran, E., Bond, N.A., Gawiser, E., Guaita, L., Feldmeier, J.J., Treister, E., Padilla, N.,
Francke, H., Matkovi\'c, A., Altmann, M., \& Herrera, D. 2012, \apj, 744, 110

\bibitem[Cowie et al.(2011)]{cowie2011} Cowie, L.~L., Barger, 
A.~J., \& Hu, E.~M.\ 2011, \apj, 738, 136

\bibitem[Dav{\'e} \etal(2011a)]{dave2011a} Dav{\'e}, R., Oppenheimer, B.D., \& 
Finlator, K. 2011, \mnras, 415, 11 

\bibitem[Dav{\'e} \etal(2011b)]{dave2011b} Dav{\'e}, R., Finlator, K., \& Oppenheimer, B.D.
2011, \mnras, 416, 1354 

\bibitem[Dawson et al.(2004)]{dawson2004} Dawson, S., Rhoads, 
J.~E., Malhotra, S., et al.\ 2004, \apj, 617, 707

\bibitem[Dawson et al.(2007)]{dawson2007} Dawson, S., Rhoads, 
J.~E., Malhotra, S., et al.\ 2007, \apj, 671, 1227 

\bibitem[Dekel \etal(2009)]{dekel2009} Dekel, A., Sari, R., \& Ceverino, D. 2009, \apj, 703, 785

\bibitem[Dijkstra \& Kramer(2012)]{dijkstra2012} Dijkstra, M., \& Kramer, R.\ 2012, 
\mnras, 424, 1672 

\bibitem[Feldmeier et al.(2002)]{feldmeier2002} Feldmeier, J.~J., 
Mihos, J.~C., Morrison, H.~L., Rodney, S.~A., 
\& Harding, P.\ 2002, \apj, 575, 779

\bibitem[Feldmeier \etal(2004)]{feldmeier2004} Feldmeier, J.J., Mihos, J.C., Morrison, H.L., 
Harding, P., Kaib, N., Dubinski, J. 2004, \apj, 609, 617 

\bibitem[Finkelstein \etal(2009a)]{fink2009} Finkelstein, S.L., Rhoads, J.E., Malhotra, S., 
\& Grogin, N. 2009, \apj, 691, 465 

\bibitem[Finkelstein et al.(2009b)]{fink2009b} Finkelstein, S.~L., 
Cohen, S.~H., Malhotra, S., \& Rhoads, J.~E.\ 2009, \apj, 700, 276 

\bibitem[Finkelstein et al.(2010)]{fink2010} Finkelstein, S.~L., 
Papovich, C., Giavalisco, M., et al.\ 2010, \apj, 719, 1250 

\bibitem[Finkelstein et al.(2011a)]{fink2011a} Finkelstein, S.~L., 
Hill, G.~J., Gebhardt, K., et al.\ 2011, \apj, 729, 140 

\bibitem[Finkelstein et al.(2011b)]{fink2011b} Finkelstein, S.~L., 
Cohen, S.~H., Moustakas, J., et al.\ 2011, \apj, 733, 117

\bibitem[Finkelstein \etal(2011c)]{fink2011c} Finkelstein, S.L., Cohen, S.H., Windhorst, R.A., 
Ryan, R.E., Hathi, N.P., Finkelstein, K.D., Anderson, J., Grogin, N.A., Koekemoer, A.M., 
Malhotra, S., Mutchler, M., Rhoads, J.E., McCarthy, P.J., O'Connell, R.W., Balick, B., 
Bond, H.E., Calzetti, D., Disney, M.J., Dopita, M.A., Frogel, J.A., Hall, D.N.B., Holtzman, J.A.,
Kimble, R.A., Luppino, G., Paresce, F., Saha, A., Silk, J.I., Trauger, J.T., Walker, A.R., 
Whitmore, B.C., \& Young, E.T. 2011, \apj, 735, 5 

\bibitem[Forero-Romero et al.(2011)]{forero2011} Forero-Romero, 
J.~E., Yepes, G., Gottl{\"o}ber, S., et al.\ 2011, \mnras, 415, 3666 

\bibitem[Francke(2009)]{francke2009} Francke, H.\ 2009, \nar, 53, 47 

\bibitem[Garel et al.(2012)]{garel2012} Garel, T., Blaizot, J., 
Guiderdoni, B., et al.\ 2012, \mnras, 422, 310

\bibitem[Gawiser \etal(2006)]{gawiser2006a} Gawiser, E., van Dokkum, P.G.,
Herrera, D., Maza, J., Castander, F.J., Infante, L., Lira, P., Quadri, R.,
Toner, R., Treister, E., Urry, C.M., Altmann, M., Assef, R., Christlein, D.,
Coppi, P.S., Dur\'an, M.F., Franx, M., Galaz, G., Huerta, L., Liu, C.,
L\'opez, S., M\'endez, R., Moore, D.C., Rubio, M., Ruiz, M.T., Toft, S., \&
Yi, S.K.  2006, \apjs, 162, 1

\bibitem[Gawiser et al.(2006)]{gawiser2006b} Gawiser, E., van 
Dokkum, P.~G., Gronwall, C., et al.\ 2006, \apjl, 642, L13 

\bibitem[Gawiser \etal(2007)]{gawiser2007} Gawiser, E., Francke, H., Lai, K.,
Schawinski, K., Gronwall, C., Ciardullo, R., Quadri, R., Orsi, A.,
Barrientos, L.F., Blanc, G.A., Fazio, G., Feldmeier, J.J., Huang, J.,
Infante, L., Lira, P., Padilla, N., Taylor, E.N., Treister, E., Urry, C.M.,
van Dokkum, P.G., \& Virani, S.N. 2007, \apj, 671, 278

\bibitem[Gonzalez et al.(2000)]{gonzalez2000} Gonzalez, A.~H., 
Zabludoff, A.~I., Zaritsky, D., \& Dalcanton, J.~J.\ 2000, \apj, 536, 561

\bibitem[Gonzalez \etal(2005)]{gonzalez2005} Gonzalez, A.H., Zabludoff, A.I., \& Zaritsky, D.
2005, \apj, 618, 195 

\bibitem[Gronwall \etal(2007)]{gronwall2007} Gronwall, C., Ciardullo, R., Hickey, T.,
Gawiser, E., Feldmeier, J.J., van Dokkum, P.G., Urry, C.M., Herrera, D.,
Lehmer, B.D., Infante, L., Orsi, A., Marchesini, D., Blanc, G.A.,
Francke, H., Lira, P., \& Treister, E. 2007, \apj, 667, 79

\bibitem[Grundahl \& Sorensen(1996)]{grundahl1996} Grundahl, F., \& Sorensen, A.~N.\ 1996, \aaps, 116, 367

\bibitem[Guaita et al.(2010)]{guaita2010} Guaita, L., Gawiser, E., 
Padilla, N.,  Francke, H., Bond, N.A., Gronwall, C., Ciardullo, R., Feldmeier, J.J.,
Sinawa, S., Blanc, G.A., \& Virani, S. 2010, \apj, 714, 255 

\bibitem[Guaita et al.(2011)]{guaita2011} Guaita, L., Acquaviva, V., Padilla, N., et al.\ 2011, \apj, 733, 114

\bibitem[Hashimoto et al.(2013)]{hashimoto2013} Hashimoto, T., Ouchi, 
M., Shimasaku, K., et al.\ 2013, \apj, 765, 70 

\bibitem[Hayashino \etal(2004)]{hayashino2004} Hayashino, T., Matsuda, Y., Tamura, H.,
Yamauchi, R., Yamada, T., Ajiki, M., Fujita, S.S., Murayama, T., Nagao, T., Ohta, K., 
Okamura, S., Ouchi, M., Shimasaku, K., Shioya, Y., \& Taniguchi, Y. 2004, \aj, 128, 2073 

\bibitem[Hibon et al.(2010)]{hibon2010} Hibon, P., Cuby, J.-G., Willis, J., et al.\ 2010, \aap, 515, A97

\bibitem[Hildebrandt \etal(2006)]{hildebrandt2006} Hildebrandt, H., 
Erben, T., Dietrich, J.P.,  Cordes, O., Haberzettl, L., Hetterscheidt, M., 
Schirmer, M., Schmithuesen, O., Schneider, P., Simon, P., \& Trachternach, C.
2006, \aap, 452, 1121 

\bibitem[Hill \etal(2007)]{hill2007} Hill, G.J.,  Gebhardt, K., Komatsu, E.,
et al. \ 2007, A.S.P. Conference Series 399, Panoramic Views of Galaxy Formation
and Evolution, ed.~T. Kodama, T. Yamada, \& K. Aoki (San Francisco:
Astronomical Society of the Pacific), 115

\bibitem[Hoversten \etal(2009)]{hoversten2009} Hoversten, E.A.,  Gronwall, C.,
Vanden Berk, D.E., Koch, T.S., Breeveld, A.A., Curran, P.A., Hinshaw, D.A.,
Marshall, F.E., Roming, P.W.A., Siegel, M.H., \& Still, M.  2009, \apj, 705, 1462

\bibitem[Hu et al.(2010)]{hu2010} Hu, E.~M., Cowie, L.~L., 
Barger, A.~J., et al.\ 2010, \apj, 725, 394

\bibitem[Jiang et al.(2013)]{jiang2013} Jiang, L., Egami, E., 
Fan, X., et al.\ 2013, arXiv:1303.0027

\bibitem[King(1971)]{king1971} King, I.R. 1971, \pasp, 83, 199 

\bibitem[Krick et al.(2006)]{krick2006} Krick, J.~E., Bernstein, 
R.~A., \& Pimbblet, K.~A.\ 2006, \aj, 131, 168 

\bibitem[Krug et al.(2012)]{krug2012} Krug, H.~B., Veilleux, S., 
Tilvi, V., et al.\ 2012, \apj, 745, 122

\bibitem[Kulas et al.(2012)]{kulas2012} Kulas, K.~R., Shapley, 
A.~E., Kollmeier, J.~A., et al.\ 2012, \apj, 745, 33 

\bibitem[Labb{\'e} et al.(2010)]{labbe2010} Labb{\'e}, I., 
Gonz{\'a}lez, V., Bouwens, R.~J., et al.\ 2010, \apjl, 716, L103

\bibitem[Lai et al.(2008)]{lai2008} Lai, K., Huang, J.-S., 
Fazio, G., et al.\ 2008, \apj, 674, 70 

\bibitem[Laursen et al.(2009)]{laursen2009} Laursen, P., 
Sommer-Larsen, J., \& Andersen, A.~C.\ 2009, \apj, 704, 1640 

\bibitem[Lehmer \etal(2005)]{lehmer2005} Lehmer, B.D., Brandt, W.N., Alexander, D.M., et al. \ 2005, \apjs, 161, 21

\bibitem[Luo \etal(2008)]{luo2008} Luo, B., Bauer, F.E., Brandt, W.N., et al. 
\ 2008, \apjs, 179, 19

\bibitem[Malhotra et al.(2012)]{malhotra2012} Malhotra, S., Rhoads, 
J.~E., Finkelstein, S.~L., et al.\ 2012, \apjl, 750, L36 

\bibitem[Mallery et al.(2012)]{mallery2012} Mallery, R.~P., 
Mobasher, B., Capak, P., et al.\ 2012, \apj, 760, 128 

\bibitem[Mart{\'{\i}}nez-Delgado et al.(2008)]{mart2008} 
Mart{\'{\i}}nez-Delgado, D., Pe{\~n}arrubia, J., Gabany, R.~J., et al.\ 
2008, \apj, 689, 184

\bibitem[Matsuda \etal(2004)]{matsuda2004} Matsuda, Y., Yamada, T., Hayashino, T., 
Tamura, H., Yamauchi, R., Ajiki, M., Fujita, S.S., Murayama, T., Nagao, T., Ohta, K., 
Okamura, S., Ouchi, M., Shimasaku, K., Shioya, Y., \& Taniguchi, Y. 2004, \aj, 128, 569 

\bibitem[Matsuda \etal(2012)]{matsuda2012} Matsuda, Y., Yamada, 
T., Hayashino, T.,  Yamauchi, R., Nakamura, Y., Morimoto, N., Ouchi, M., Ono, Y., 
Umemura, M., \& Mori, M. 2012, \mnras, 425, 878

\bibitem[McLinden et al.(2011)]{mclinden2011} McLinden, E.~M., 
Finkelstein, S.~L., Rhoads, J.~E., et al.\ 2011, \apj, 730, 136

\bibitem[Melnick et al.(1999)]{melnick1999} Melnick, J., Selman, 
F., \& Quintana, H.\ 1999, \pasp, 111, 1444

\bibitem[Morrison \etal(1994)]{morrison1994} Morrison, H.L., 
Boroson, T.A., \& Harding, P. 1994, \aj, 108, 1191 

\bibitem[Murayama et al.(2007)]{murayama2007} Murayama, T., 
Taniguchi, Y., Scoville, N.~Z., et al.\ 2007, \apjs, 172, 523

\bibitem[Muller \etal(1998)]{muller1998} Muller, G.P., Reed, R., 
Armandroff, T., Boroson, T.A., \& Jacoby, G.H. 1998, \procspie, 3355, 577 

\bibitem[Nakajima et al.(2012)]{nakajima2012} Nakajima, K., Ouchi, 
M., Shimasaku, K., et al.\ 2012, \apj, 745, 12 

\bibitem[Neufeld(1991)]{neufeld1991} Neufeld, D.A.\ 1991, \apjl, 370, L85 

\bibitem[Nilsson et al.(2007)]{nilsson2007} Nilsson, K.~K., M{\o}ller, P., M{\"o}ller, O., et al.\ 2007, \aap, 471, 71

\bibitem[Nilsson et al.(2009)]{nilsson2009} Nilsson, K.~K., Tapken, C., M{\o}ller, P., et al.\ 2009, \aap, 498, 13

\bibitem[Nilsson et al.(2011)]{nilsson2011} Nilsson, K.~K., {\"O}stlin, G., M{\o}ller, P., et al.\ 2011, \aap, 529, A9

\bibitem[Ono et al.(2010)]{ono2010} Ono, Y., Ouchi, M., 
Shimasaku, K., et al.\ 2010, \mnras, 402, 1580

\bibitem[Orsi et al.(2012)]{orsi2012} Orsi, A., Lacey, C.~G., \& Baugh, C.~M.\ 2012, \mnras, 425, 87

\bibitem[{\"O}stlin et al.(2009)]{ostlin2009} {\"O}stlin, G., 
Hayes, M., Kunth, D., et al.\ 2009, \aj, 138, 923 

\bibitem[Ota et al.(2010)]{ota2010} Ota, K., Iye, M., 
Kashikawa, N., et al.\ 2010, \apj, 722, 803 

\bibitem[Oteo et al.(2012)]{oteo2012} Oteo, I., Bongiovanni, A., 
P{\'e}rez Garc{\'{\i}}a, A.~M., et al.\ 2012, \apj, 751, 139 

\bibitem[Ouchi et al.(2008)]{ouchi2008} Ouchi, M., Shimasaku, K., 
Akiyama, M., et al.\ 2008, \apjs, 176, 301 

\bibitem[Ouchi et al.(2010)]{ouchi2010} Ouchi, M., Shimasaku, K., 
Furusawa, H., et al.\ 2010, \apj, 723, 869

\bibitem[Paudel et al.(2013)]{paudel2013} Paudel, S., Duc, P.-A., 
C{\^o}t{\'e}, P., et al.\ 2013, \apj, 767, 133

\bibitem[Peng \etal(2002)]{galfit2002} Peng, C.Y., Ho, L.C., Impey, C.D., \& Rix, H. 
2002, \aj, 124, 266

\bibitem[Peng \etal(2010)]{galfit2010} Peng, C.Y., Ho, L.C., Impey, C.D., \& Rix, H.-W. 
2010, \aj, 139, 2097 

\bibitem[Pirzkal \etal(2007)]{pirzkal2007} Pirzkal, N., Malhotra, S., Rhoads, J.E., \& Xu, C.
2007, \apj, 667, 49

\bibitem[Prescott et al.(2012)]{prescott2012} Prescott, M.~K.~M., 
Dey, A., \& Jannuzi, B.~T.\ 2012, \apj, 748, 125

\bibitem[Press et al.(1992)]{press1992} Press, W.~H., Teukolsky, 
S.~A., Vetterling, W.~T., \& Flannery, B.~P.\ 1992, Numerical Recipes in Fortran -- The
Art of Scientific Computing, 2nd ed. (Cambridge: University Press).  

\bibitem[Rauch \etal(2008)]{rauch2008} Rauch, M., Haehnelt, M., 
Bunker, A., Becker, G., Marleau, F., Graham, J., Cristiani, S., Jarvis, M., Lacey, C., 
Morris, S., Peroux, C., R\"ottgering, H., \& Theuns, T. 2008, \apj, 681, 856

\bibitem[Rudick \etal(2010)]{rudick2010} Rudick, C.S., Mihos, J.C., Harding, P., 
Feldmeier, J.J., Janowiecki, S., \& Morrison, H.L. 2010, \apj, 720, 569 

\bibitem[Saito et al.(2006)]{saito2006} Saito, T., Shimasaku, K., 
Okamura, S., et al.\ 2006, \apj, 648, 54

\bibitem[Salvadori et al.(2010)]{salva2010} Salvadori, S., Dayal, 
P., \& Ferrara, A.\ 2010, \mnras, 407, L1 

\bibitem[Schechter(1976)]{schechter1976} Schechter, P. 1976, \apj, 203, 297

\bibitem[Schiminovich \etal(2003)]{schim2003} Schiminovich, D., Arnouts, S., Milliard, B., 
\& GALEX Science Team 2003, Bulletin of the American Astronomical Society, 35, \#103.04 

\bibitem[Shapley \etal(2001)]{shapley2001} Shapley, A.E., Steidel, C.C., Adelberger, K.L., 
Dickinson, M., Giavalisco, M., \& Pettini, M. 2001, \apj, 562, 95

\bibitem[Shapley \etal(2003)]{shapley2003} Shapley, A.E., Steidel, C.C., Pettini, M., 
\& Adelberger, K.L. 2003, \apj, 588, 65 

\bibitem[Shimasaku et al.(2004)]{shim2004} Shimasaku, K., 
Hayashino, T., Matsuda, Y., et al.\ 2004, \apjl, 605, L93 

\bibitem[Shimasaku et al.(2006)]{shim2006} Shimasaku, K., 
Kashikawa, N., Doi, M., et al.\ 2006, \pasj, 58, 313 

\bibitem[Slater \etal(2009)]{slater2009} Slater, C.T., Harding, P., \& Mihos, J.C. 2009, 
\pasp, 121, 1267 

\bibitem[Smith et al.(2012)]{smith2012} Smith, B.~M., Malhotra, 
S., Rhoads, J., et al.\ 2012, American Astronomical Society Meeting 
Abstracts \#219, 219, \#340.06

\bibitem[Stark et al.(2009)]{stark2009} Stark, D.~P., Ellis, 
R.~S., Bunker, A., et al.\ 2009, \apj, 697, 1493

\bibitem[Steidel \etal(1996)]{steidel1996} Steidel, C.C., Giavalisco, M., Pettini, M., 
Dickinson, M., \& Adelberger, K.L. 1996, \apjl, 462, L17 

\bibitem[Steidel \etal(1998)]{steidel1998} Steidel, C.C., Adelberger, K.L., Dickinson, M., 
Giavalisco, M., Pettini, M., \& Kellogg, M. 1998, \apj, 492, 428

\bibitem[Steidel \etal(2000)]{steidel2000} Steidel, C.C., Adelberger, K.L., Shapley, A.E., 
Pettini, M., Dickinson, M., \& Giavalisco, M. 2000, \apj, 532, 170 

\bibitem[Steidel \etal(2010)]{steidel2010} Steidel, C.C., Erb, D.K., Shapley, A.E., 
Pettini, M., Reddy, N., Bogosavljevi\'c, M., Rudie, G.C., \& Rakic, O. 2010, \apj, 717, 289 

\bibitem[Steidel \etal(2011)]{steidel2011} Steidel, C.C. Bogosavljevi\'c, M., Shapley, A.E.,
Kollmeier, J.A., Reddy, N.A., Erb, D.K., \& Pettini, M. 2011, \apj, 736, 160 (S11)

\bibitem[Stetson(1987)]{stetson1987} Stetson, P.B.\ 1987, \pasp, 99, 191 

\bibitem[Taniguchi et al.(2005)]{taniguchi2005} Taniguchi, Y., Ajiki, 
M., Nagao, T., et al.\ 2005, \pasj, 57, 165 

\bibitem[Tapken et al.(2006)]{tapken2006} Tapken, C., Appenzeller, I., Gabasch, A., et al.\ 2006, \aap, 455, 145

\bibitem[Tapken et al.(2007)]{tapken2007} Tapken, C., Appenzeller, I., Noll, S., et al.\ 2007, \aap, 467, 63

\bibitem[Tilvi et al.(2010)]{tilvi2010} Tilvi, V., Rhoads, J.~E., Hibon, P., et al.\ 2010, \apj, 721, 1853 

\bibitem[van Breukelen et al.(2005)]{van2005} van Breukelen, 
C., Jarvis, M.~J., \& Venemans, B.~P.\ 2005, \mnras, 359, 895 

\bibitem[Veilleux \etal(2005)]{veilleux2005} Veilleux, S., Cecil, G., \& Bland-Hawthorn, J.
2005, \araa, 43, 769

\bibitem[Venemans et al.(2004)]{venemans2004} Venemans, B.~P., R{\"o}ttgering, H.~J.~A., Overzier, R.~A., et al.\ 2004, \aap, 424, L17 

\bibitem[Verhamme et  al.(2006)]{verhamme2006} Verhamme, A., Schaerer, D., \& Maselli, A.\ 2006, \aap, 460, 397 

\bibitem[Verhamme et al.(2008)]{verhamme2008} Verhamme, A., Schaerer, D., Atek, H., \& Tapken, C.\ 2008, \aap, 491, 89 

\bibitem[Verhamme et al.(2012)]{verhamme2012} Verhamme, A., Dubois, Y., Blaizot, J., et al.\ 2012, \aap, 546, A111

\bibitem[Virani \etal(2006)]{virani2006} Virani, S.N., Treister, E., Urry, C.M., \& Gawiser, E.\
2006, \aj, 131, 2373

\bibitem[Yuma et al.(2010)]{yuma2010} Yuma, S., Ohta, K., Yabe, 
K., et al.\ 2010, \apj, 720, 1016

\bibitem[Wang et al.(2005)]{wang2005} Wang, J.~X., Malhotra, S., 
\& Rhoads, J.~E.\ 2005, \apjl, 622, L77 

\bibitem[Wang et al.(2009)]{wang2009} Wang, J.-X., Malhotra, S., 
Rhoads, J.~E., Zhang, H.-T., \& Finkelstein, S.~L.\ 2009, \apj, 706, 762 

\bibitem[Westra et al.(2006)]{westra2006} Westra, E., Jones, D.~H., Lidman, C.~E., et al.\ 2006, \aap, 455, 61 

\bibitem[Windhorst \etal(2011)]{windhorst2011} Windhorst, R.A., Cohen, S.H., Hathi, N.P., et al. 
2011, \apjs, 193, 27

\bibitem[Yajima et al.(2012)]{yajima2012} Yajima, H., Li, Y., Zhu, 
Q., et al.\ 2012, \apj, 754, 118 

\bibitem[Zheng \etal(1999)]{zheng1999} Zheng, Z., Shang, Z., Su, H.,
et al. 1999, \aj, 117, 2757 


\bibitem[Zheng \etal(2011)]{zheng2011} Zheng, Z., Cen, R., Weinberg, D., Trac, H., \& 
Miralda-Escud{\'e}, J. 2011, \apj, 739, 62 

\bibitem[Zibetti \etal(2005)]{zibetti2005} Zibetti, S., White, S.D.M., Schneider, D.P., 
\& Brinkmann, J. 2005, \mnras, 358, 949 

\end{thebibliography}
\end{document}